\documentclass[usenatbib]{mn2e}
\usepackage{graphicx}
\usepackage{abbreviations}
\usepackage{natbib}
\usepackage{amssymb}

\newcommand{\be}{\begin{equation}}
\newcommand{\ee}{\end{equation}}
\newcommand{\nn}{\mbox{} \nonumber \\ \mbox{} }
\newcommand{\ba}{\begin{eqnarray}}
\newcommand{\ea}{\end{eqnarray}}

\newcommand\eg{\textit{e.g.,\ }}

\newcommand{\fracp}[2]{\left(\frac{#1}{#2}\right)}

\newcommand{\ort}[1]{{\bf n}_{#1}}

\begin{document}

\title[The Inner Knot]
{The Inner Knot of The Crab Nebula}

\author[Lyutikov et al.]
{
Maxim Lyutikov$^{1}$\thanks{E-mail: lyutikov@purdue.edu}
Serguei S. Komissarov$^{1,2}$\thanks{E-mail: s.s.komissarov@leeds.ac.uk (SSK)},
Oliver Porth$^{2}$\thanks{E-mail: o.porth@leeds.ac.uk (OP)}\\
$^{1}$Department of Physics and Astronomy, Purdue University, West Lafayette, IN 47907-2036, USA\\
$^{2}$Department of Applied Mathematics, The University of Leeds, Leeds, LS2 9JT, UK 
}

\date{Received/Accepted}
\maketitle

\begin{abstract}
We model the inner knot of the Crab Nebula as a synchrotron emission 
coming from the non-spherical MHD termination shock of relativistic pulsar wind.
The post-shock flow is mildly relativistic; as a result   the  
Doppler-beaming has a strong impact on the shock appearance.   
The  model can reproduce  the knot location, size, elongation, 
brightness distribution, luminosity and polarization provided the effective 
magnetization  of the section of the pulsar wind producing the  knot  is low, $\sigma \leq 1$.
In the striped wind model, this implies that the striped zone is rather wide, 
with the magnetic inclination angle of the Crab pulsar $\ge 45^\circ$; this agrees with 
the  previous model-dependent estimate based on the gamma-ray emission of the 
pulsar. We conclude that the tiny knot is indeed a bright spot on 
the surface of a quasi-stationary magnetic relativistic shock and that this
shock is a site of efficient particle acceleration. On the other hand,  
the deduced low magnetization of the knot plasma implies that this is an 
unlikely site for the Crab's gamma-ray flares, if they are related to the 
fast relativistic magnetic reconnection events.
\end{abstract}

\begin{keywords}
ISM: supernova remnants -- MHD -- shock waves -- gamma-rays: theory --
radiation mechanisms: non-thermal -- relativity -- pulsars: individual: Crab
\end{keywords}

\section{Introduction}

The Crab pulsar and its pulsar wind nebula (PWN) remain prime targets for 
high energy astrophysical research. In many ways, the current models
of Active Galactic Nuclei and Gamma Ray Bursts are based on what we have 
learned from the studies of the Crab.  The  recent detection of flares
from the Crab Nebula by AGILE and Fermi satellites
\citep{2011Sci...331..736T,2011Sci...331..739A} have brought this object 
into the ``focal point'' once again.  Their extreme properties seem impossible
to explain within the standard theories of non-thermal particle acceleration 
and require their overhaul with important implications to 
high energy astrophysics in general 
\citep[e.g.][]{2010MNRAS.405.1809L,2012MNRAS.426.1374C,2012ApJ...746..148C,2012MNRAS.427.1497L,2014RPPh...77f6901B}

In the MHD models of the Crab Nebula, the super-fast-magnetosonic
relativistic wind of the Crab pulsar terminates at a reverse shock
\citep{reesgunn,kc84}.  
However, finding the shock in the images of the Crab Nebula has 
not been a straight-forward matter - there seem to be no sharp feature which can 
be undoubtedly identified with the shock surface. In their seminal paper,  
\citet{kc84} discuss the under-luminous region hosting the Crab pulsar and surrounded 
by the optical wisps as an indicator of the shock presence. After the discovery of the 
inner X-ray ring by Chandra 
\citep{2000ApJ...536L..81W,2002ApJ...577L..49H}, the ring is often referred to as 
the termination shock and yet 
this feature looks much more like a collection of knots than a smooth surface.  
A new twist in the story has come with the recent PIC simulations which show 
the shock particle acceleration is highly inefficient in even relatively weakly 
magnetized relativistic plasma \citep{SironiSpitkovsky09,
2011ApJ...726...75S}. These results make one doubt that the shock can be visible 
at all.     
On the other hand, the wind from an oblique rotator should have the so-called 
striped zone where the orientation of magnetic field alternated on the scale of 
the pulsar period.  The magnetic energy associated with these stripes can 
be dissipated at the termination shock and converted into the energy of the wind 
particles \citep{2003MNRAS.345..153L,2011ApJ...741...39S,2011ApJ...741...39S,
2013ApJ...770...18A,2007A&A...473..683P}.

Given the highly anisotropic nature of the
wind, the termination shock is squashed along the polar direction and can be
highly oblique with respect to the upstream flow
\citep{2002MNRAS.329L..34L}. Downstream of the shock, the flow can
still be relativistic and its emission subject to strong
Doppler beaming. The computer simulations of the Crab nebula and its
radiation \citep{KomissarovLyubarsky} revealed the presence of a very
bright compact feature in the synthetic synchrotron maps, highly
reminiscent of the HST knot 1 of the Crab Nebula located very close
to the pulsar \citep[also called the inner knot, ][]{1995ApJ...448..240H}.  
(In these simulations, the termination shock was treated as source of 
synchrotron electrons with power-law energy spectrum, which then were carried out 
into the nebula by the shocked wind plasma.)
This feature, confirmed in the later
more advanced 2D \citep{2009MNRAS.400.1241C} and 3D
\citep{2014MNRAS.438..278P} simulations, is associated with the
location at the termination shock where the shocked plasma flows in
the direction of the fiducial observer and thus strongly
Doppler-boosted.  \cite{2011MNRAS.414.2017K} argued that given the
short synchrotron life-time of the high energy electrons compared to
the dynamical time-scale of the shock, the knot can be the main
source of the gamma-ray emission from the Nebula at 10-100 MeV.

Recently, a targeted multi-wavelength study of the Crab's inner knot
has been conducted by \cite{2015arXiv150404613R} in order to check if it shows any 
activity correlated with the gamma-ray flares. 
Although no such correlation has been found, the optical 
data reveal the structure and temporal evolution of the knot with unprecedented 
detail. In this paper, we investigate if the data are consistent with the MHD-shock model of 
the knot using simple analytical and semi-analytical tools.  In particular, we combine 
the theoretical shape of the shock with the oblique shock jumps in order to obtain 
the Doppler-beaming of the post shock emission and use this to determine the location, 
the shape and the brightness distribution of the knot. 


\section{Geometry of the termination shock}
\label{Shock-Shape}

At the location of the termination shock, the magnetic field of the 
pulsar wind has the form of loops centered on the pulsar's rotational axis. 
The wind's termination shock is also symmetric with respect to the rotational 
axis and hence the magnetic field is parallel to the shock surface. 
The pulsar wind is not spherical -- its luminosity per unit solid angle increases with 
the polar angle measured from the pulsar rotational axis.
As a result, the termination shock is not spherical and the radial stream lines 
of the wind are generally not normal to the shock surface -- locally the shock is oblique.  
In addition, the pulsar wind is ultra-relativistic and its thermal pressure is
negligibly small. The corresponding shock 
equations have been analyzed in \cite{2011MNRAS.414.2017K,2012MNRAS.422.3118L}; 
see also Appendix \ref{oblique}. Here we summarize their results using the 
notation introduced in \cite{2011MNRAS.414.2017K}.

We differentiate the flow parameters upstream and downstream of the shock 
using indices ``1'' and ``2'' respectively. 
Denote as $\delta$ the angle between the velocity vector and the shock 
surface (the angle of attack). Then in the observer's frame 
\be
     \tan\delta_2= \chi\tan\delta_1
\ee
and for the Lorentz factor of the flow
\be
     \Gamma_2= \Gamma_1\left[1+\Gamma_1^2\sin^2\delta_1(1-\chi^2) \right]^{-1/2}\,, 
\ee
where $\chi=v_{n2}/v_{n1}$ is the ratio of the normal velocity components. 
For a strong shock, $\Gamma_2 \ll \Gamma_1$ and the last equation reduces to 
\be
     \Gamma_2= (1-\chi^2)^{-1/2}\csc\delta_1\,. 
\label{gamma2}
\ee
Assuming $\delta_1 \gg 1/\Gamma_1$ and using the ratio of specific heats $\gamma=4/3$, 
\cite{2011MNRAS.414.2017K} obtained  
\begin{equation}
\chi = \frac{1+2\sigma_1 + \sqrt{16\sigma_1^2+16\sigma_1+1}}{6(1+\sigma_1)}\,, 
\label{chi1}
\end{equation}
where $\sigma_1=B_1^{'2}/\rho'_1$ is the magnetization parameter of the wind, $B'$ and 
$\rho'$ are the comoving values of the magnetic field and the rest-mass density of
plasma respectively.  
This is a monotonic function increasing from $\chi(0)=1/3$ to $\chi(+\infty)=1$.  
For $\sigma_1\gg1$, one can use the approximation 
\begin{equation}
\chi \simeq  1 -\frac{1}{2\sigma_1}\,. 
\label{chi2}
\end{equation}

Using Eq. (\ref{gamma2}) we find that for $\sigma_1=0$ 
\begin{equation}
   \Gamma_2 = \frac{3}{2\sqrt{2}}\csc\delta_1\,
\label{gam-low}
\end{equation}
and for $\sigma_1\gg1$
\begin{equation}
   \Gamma_2 \simeq \sqrt{\sigma_1}\csc\delta_1\,. 
\label{gam-high}
\end{equation}

The deflection $\Delta\delta = \delta_1-\delta_2$ is given by 
\be
    \tan\Delta\delta = \frac{\tan\delta_1(1-\chi)}{1+\chi\tan^2\delta_1}\,.
\label{def-angle}
\ee  
It reaches the maximum value of 
\be 
     \tan(\Delta\delta_{max}) = \frac{1}{2}\frac{1-\chi}{\sqrt{\chi}}
\quad\mbox{at}\quad   \tan\delta_{1} = \chi^{-1/2} \,.
\ee
For $\sigma_1=0$ this gives $\Delta\delta_{max} = \pi/6$ at  $\delta_1= \pi/3$, 
whereas for $\sigma_1\gg1$ one has  $\Delta\delta_{max} = 1/4\sigma_1$ at $\delta_1 = \pi/4$.

The total pressure $\tilde{p}_{2} = p_2 +\frac{B_2^2}{2}$ downstream of the shock is 
\begin{equation}
\tilde{p}_2 = (1-\chi) (F/c) \sin^2\delta_1 \,, 
\label{pt2am}
\end{equation}
where $F$ is the upstream total energy flux density along the flow velocity 
(see Appendix~\ref{oblique}).  For $\sigma_1=0$, this yields
\begin{equation}
\tilde{p}_2 = \frac{2}{3}\frac{F}{c} \sin^2\delta_1\,  
\label{pt2b}
\end{equation}
whereas for $\sigma_1\gg1$, 
\begin{equation}
\tilde{p}_2 = \frac{1}{2\sigma_1}\frac{F}{c} \sin^2\delta_1\,.  
\label{pt2c}
\end{equation}
One can see that for the same energy flux the post-shock pressure is significantly
reduced compared to the purely hydro case.

Since the shock is driven into the wind by the pressure inside the nebula, $p_n$, which 
is approximately uniform in the nebula due to its slow expansion, we replace 
$\tilde{p}_2$ with constant $p_n$, which makes our approach similar to the Kompaneets 
approximation \citep{Komp}.  This approximation was already used by \cite{2002MNRAS.329L..34L}, 
to determine the shape of the termination shock for a weakly magnetized wind. 
It less clear if the approximation can hold well for the polar section of the shock
where the magnetization and the Lorentz factor of the postshock flow can be very high. 
This makes terms other than the total pressure potentially important 
in the transverse force balance. This is already seen in the numerical simulations with 
moderate wind magnetization, where the magnetic hoop stress leads to compression 
of the polar region \citep{2014MNRAS.438..278P}. Moreover, these simulations show that 
the polar flow is highly variable. Keeping these in mind, we shell still shell proceed 
exploring the models based on the assumptiopn $\tilde{p}_2=p_n=$ const.

\begin{figure}
\includegraphics[width=\columnwidth]{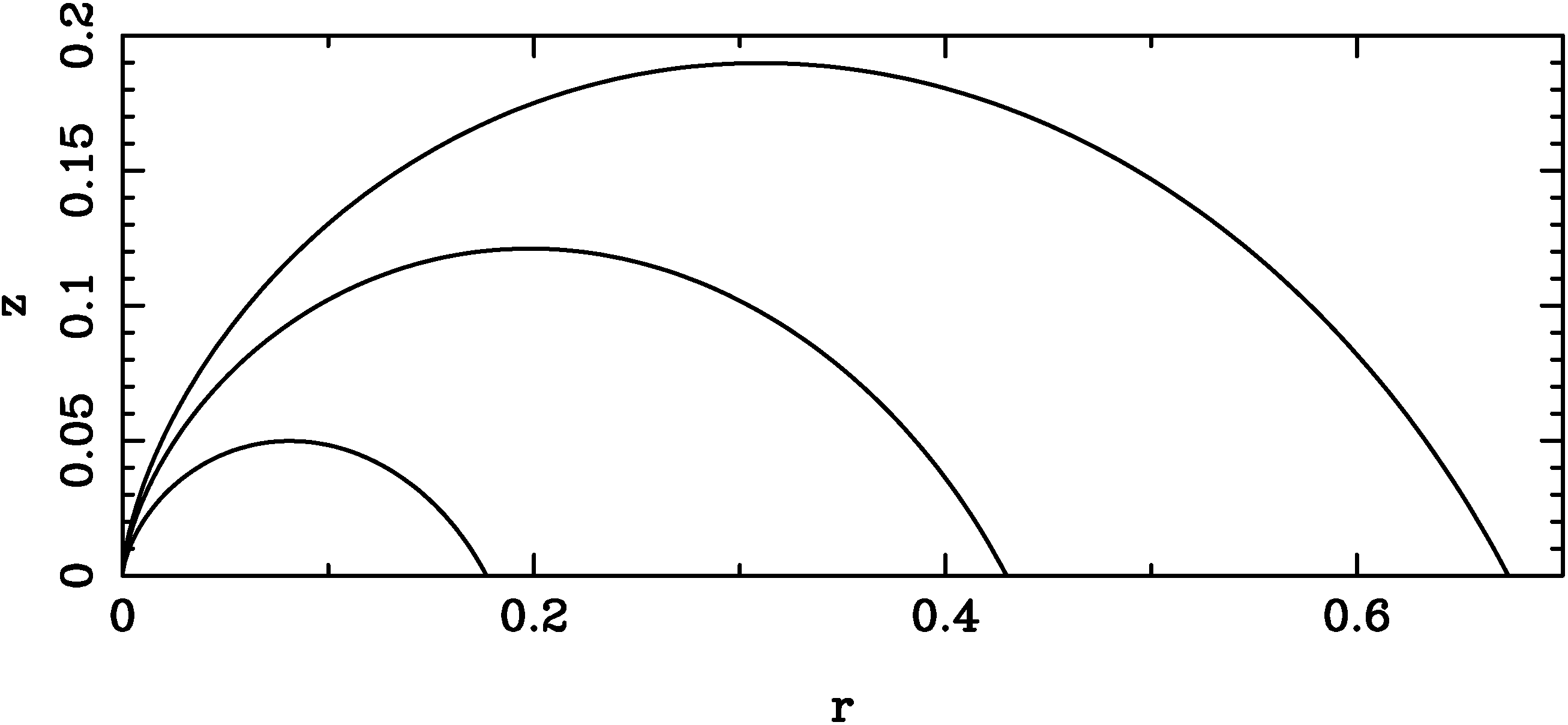}
\includegraphics[width=\columnwidth]{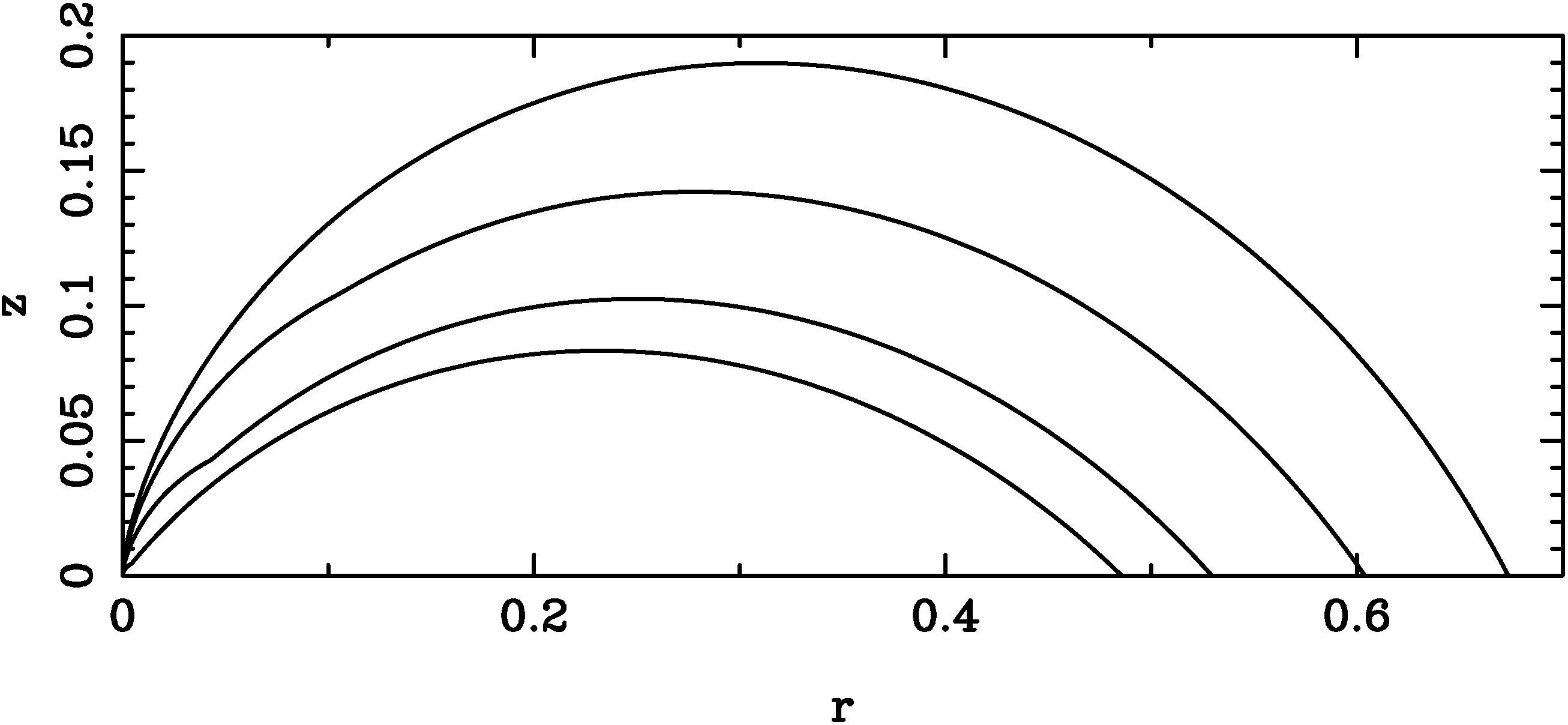}
\includegraphics[width=\columnwidth]{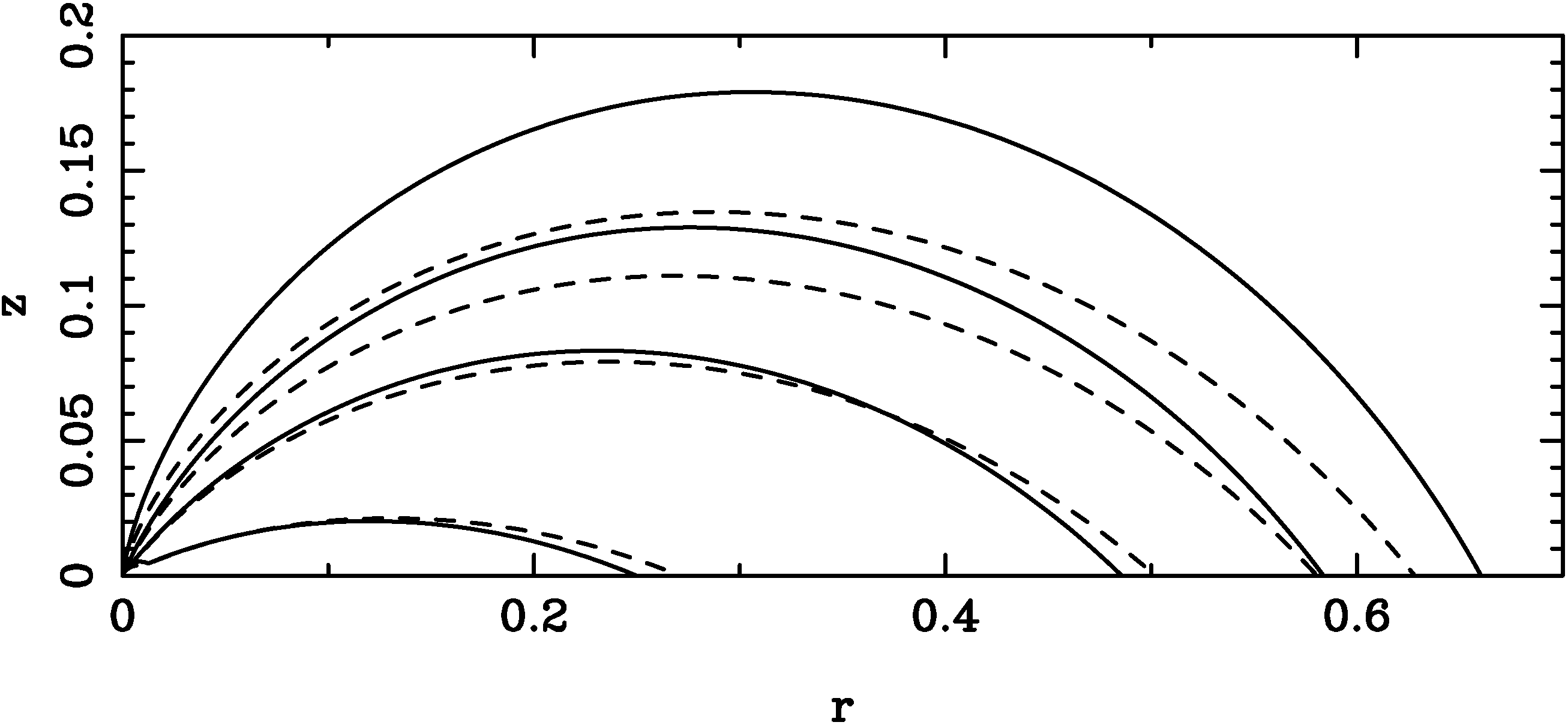}
\caption{
{\it Top panel}: 
Shape of the termination shock for a uniform distribution of $\sigma_1$. 
The curves correspond to $\sigma_1=0,1$ and 10, from top to bottom.
{\it Middle panel}: 
Shape of the termination shock for the non-uniform distribution of $\sigma_1$ 
corresponding to a striped wind with $\alpha=\pi/4$ and $f(\theta)=\sin^2\theta$.
The lines correspond to $\sigma_0 = 0,1,10$ and $10^3$, from top to bottom.
{\it Bottom panel}: 
Shape of the termination shock for the non-uniform distribution of $\sigma_1$
corresponding to a striped wind  $\sigma_0=10^3$ and $f(\theta)=\sin^n\theta$.
The lines correspond to $\alpha = 80^o,60^o,45^o$ and $20^o$, from top to bottom. 
The solid lines correspond to $n=2$ and the dashed lines to $n=4$. 
} 
\label{fig:all-shapes}
\end{figure}

If the function $R(\theta)$ gives the spherical radius as a function of the polar angle
on the shock surface then 
\begin{equation}
    \tan\delta_1=\fracp{R}{R'}\,.
\label{sin}
\end{equation}
For an axisymmetric radial wind, its energy flux can be written as 
$F=L_0 f(\theta)/4\pi R^2$, where $f(\theta)$ describes the wind anisotropy. 
We will consider only $f=\sin^n\theta$, where $n=2$ for the monopole model of the 
pulsar magnetosphere \citep{1999A&A...349.1017B}. 
Recently, \citet{2015ApJ...801L..19P}  argued for $n=4$, based on their numerical 
simulations of pulsar magnetospheres. 
Substituting the expressions for $\sin\delta_1$ and $F$ into Eq. (\ref{pt2am}), we 
obtain the shock-shape equation 
\begin{equation}
    R'^2+R^2 = \frac{L_0f(\theta)}{4\pi c p_n} (1-\chi) \,.
\label{shock-shape-eq}
\end{equation}
Finally, we introduce the characteristic length scale of the problem 
$R_0=L_0f(\theta)/4\pi c p_n$ and arrive to the dimensionless equation 
\begin{equation}
    X'^2+X^2 = f(\theta)(1-\chi) \,,
\label{shock-shape}
\end{equation}
where $X=R/R_0$. (This is the modified version of our original Eq.3.) 
The appropriate boundary condition is 
\begin{equation}
X(0)=0 \,.  
\end{equation}

\begin{figure}
\centering
\includegraphics[width=0.8\linewidth]{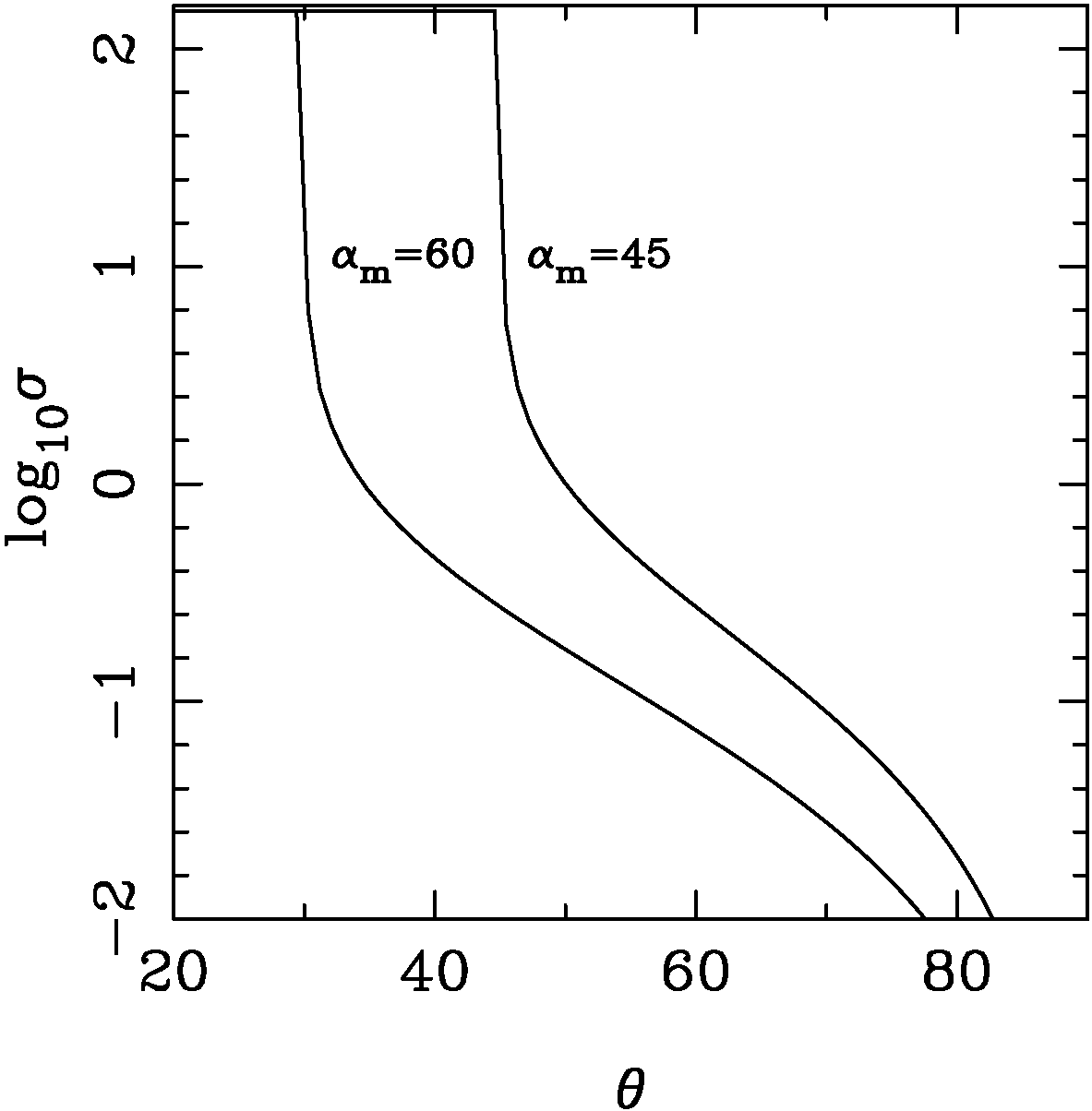}
\caption{Magnetization of the striped wind zone after the dissipation of 
stripes for the magnetic inclination angles $\alpha_m=60$ and 45 degrees. }
\label{striped-sigma}
\end{figure}

When the shock terminates the striped part of the pulsar wind, the shock solution 
is modified due to the dissipation of the magnetic energy associated with the stripes. 
\citet{2003MNRAS.345..153L} have shown that the shock solution is actually the same as that for 
the unstriped flow where the energy of stripes is already converted into the bulk 
kinetic energy of the wind particles. Thus, as long as the shock solution is concerned 
it does not matter where the dissipation occurs, in the wind or at the shock.       
The magnetization of the wind that has lost its stripes can be found as 
\begin{equation}
    \sigma_1=\sigma_0 \frac{\chi_\alpha(\theta)}{1+\sigma_0(1-\chi_\alpha(\theta))}\,,
\label{eff-sigma}
\end{equation}
where
\begin{equation}
\chi_\alpha(\theta) = \left\{
    \begin{array}{ll}
       (2\phi_\alpha(\theta)/\pi-1)^2, &
      \theta_m<\theta<\pi/2\\ 1, & \theta\le\theta_m
\end{array}
\right. ,
\end{equation}
and
$$
  \cos\phi_\alpha = \frac{\tan\theta_m}{\tan\theta}\,
$$   
\citep{2013MNRAS.428.2459K}.
In these equations, $\theta_m=\pi/2-\alpha$, where $\alpha$ is the pulsar's magnetic 
inclination angle, is the polar angle of the boundary separating the unstriped polar 
section of the pulsar wind from its equatorial striped zone and $\sigma_0$ is the original 
magnetization of the striped wind. Figure~\ref{striped-sigma} shows the wind magnetization 
after the dissipation of its stripes for $\sigma_0=100$ and $\alpha_m=60^o$ and 
$45^o$ degrees. The most interesting feature of these solutions is the rapid drop of 
$\sigma$ at the boundary of the striped zone.

Eq. (\ref{shock-shape}) is integrated numerically. Due to its singularity at $\theta=0$, 
we use its asymptotic analytic solution
$$
    X=\frac{2(1-\chi(\sigma_0)) }{n+2} \theta^{\frac{n}{2}+1}  
$$
in order to move away from the origin.

As a start, we consider the case of uniform $\sigma_1$, where it does not depend on the 
polar angle. This corresponds to the case of aligned rotator, $\alpha=0$, where 
$\sigma_1=\sigma_0$ everywhere. The top panel of 
Figure~\ref{fig:all-shapes} shows the solutions for $\sigma_1=0,1$ and 10. 
As one can see, for higher $\sigma_1$ the shock is located closer to the pulsar. 
This is in agreement with the earlier results by \citet{kc84}. In fact, the curves 
differ only by the scaling factor $\sqrt{1-\chi(\sigma_1)}$, as follows from 
Eq. (\ref{shock-shape}).    
 
When the variation of $\sigma_1$ due to the existence of the striped 
wind zone (see Eq. (\ref{eff-sigma})) is taken into account, the variation of the 
shock size is less dramatic. In the middle panel of Figure~\ref{fig:all-shapes}, 
we show the solutions for $\alpha=\pi/4$ and $f(\theta)=\sin^2\theta$, corresponding to 
different values of the magnetization parameter $\sigma_0$. 
As $\sigma_0$ increases, the shock still becomes 
more compact, but as $\sigma_0\to\infty$ the dependence  becomes very weak and the 
shock approaches some asymptotic shape. Such a turn is clearly connected with 
the existence of the striped wind section where $\sigma_1(\theta)$ becomes 
insensitive to $\sigma_0$:        
\begin{equation}
    \sigma_1 \simeq \frac{\chi_\alpha(\theta)}{(1-\chi_\alpha(\theta))}\,.
\label{eff-sigma-ass}
\end{equation}

The bottom panel of Figure~\ref{fig:all-shapes} illustrates the dependence of the 
shock shape on the magnetic inclination angle $\alpha$ for $\sigma_0=10^3$ and 
$f(\theta)=\sin^n\theta$ with $n=2,4$. 
As one can see, the shock becomes more compact as $\alpha$ decreases. 
This is expected, as for $\alpha=0$, the case of uniform magnetization with 
$\sigma_1=10^3$ is recovered and in this case the shock size rapidly decreases 
with $\sigma_0$.  However, even for $\alpha=20^o$ the equatorial radius 
of the shock is still much larger than that in the limiting case of $\alpha=0$.  
For n=2, the total 
wind power is $L_w=(10/15)L_0$, whereas for n=4 it is $L_w=(8/15)L_0$. Since it is more 
interesting to compare the results corresponding to the same wind power, we re-scale 
the n=4 solution of Eq. (\ref{shock-shape}) by the factor $\sqrt{5}/2$.  
In the bottom panel of Figure~\ref{fig:all-shapes} the n=2 solutions 
are shown as  solid lines and the n=4 solutions as dashed lines. 
The difference between the two groups is not large, particularly for $\alpha\le 45^o$.

\begin{figure}
\includegraphics[width=7cm]{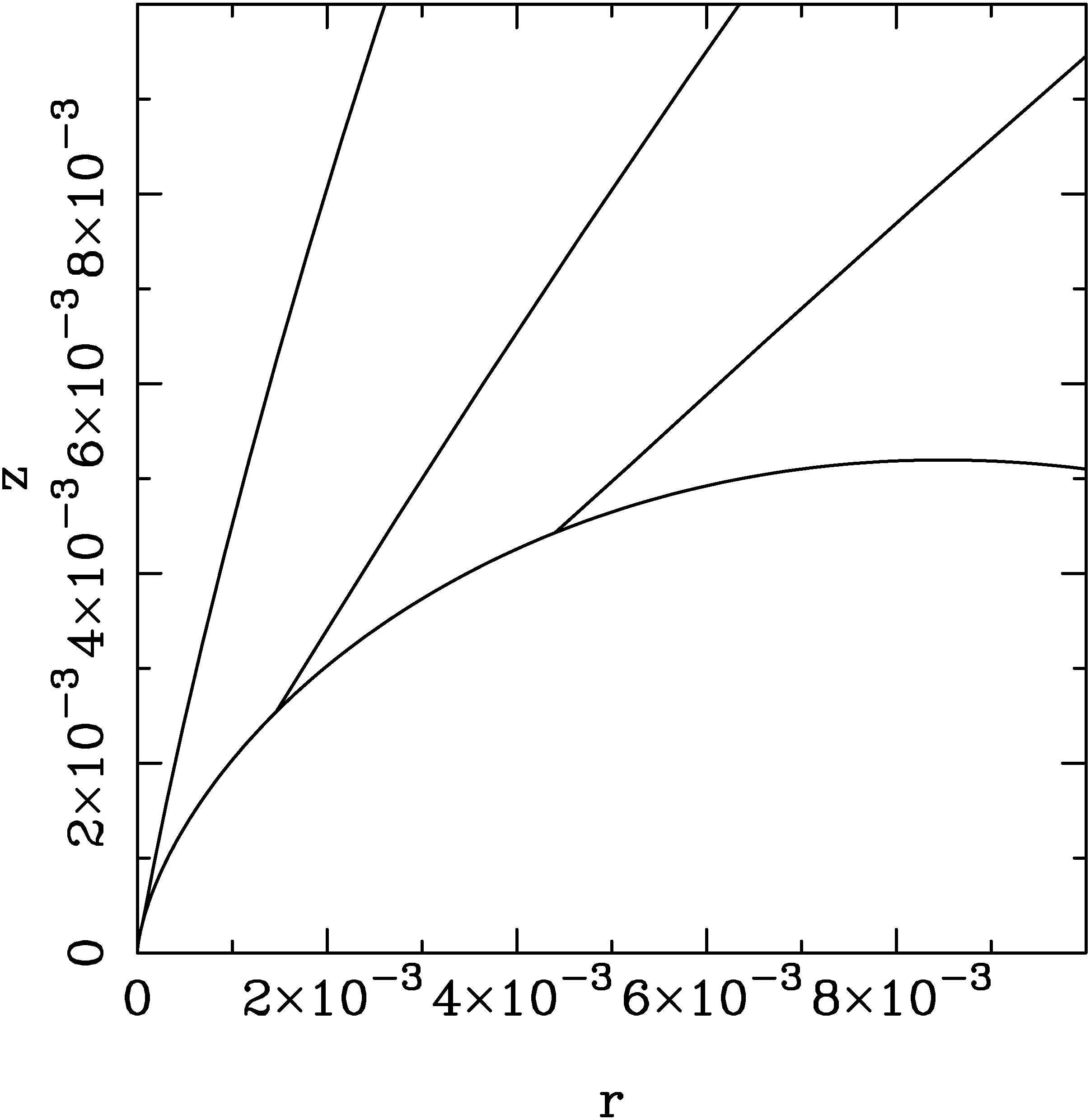}
\caption{Shape of the termination shock near the origin. 
The solutions correspond to the model with $\sigma_0=10^3$ and $f(\theta)=\sin^2\theta$.
As in the middle panel of Figure~\ref{fig:all-shapes}, the magnetic inclination angle 
$\alpha = 80^o,60^o,45^o$ and $20^o$, from top to bottom.
} 
\label{fig:n2b}
\end{figure}

Figure~\ref{fig:n2b} zooms into the inner region of the middle panel of 
Figure \ref{fig:all-shapes}, where the shock exhibits a noticeable break.  
The origin of this break is easy to understand. At $\theta<\theta_m$, the magnetization 
$\sigma_1=\sigma_0$ is constant. Hence, the shock curve is a miniature version of 
that of pure hydro shock (see Eqs.\ref{pt2b} and \ref{pt2c}). At $\theta=\theta_m$, 
$\sigma_1(\theta)$ rapidly drops leading to higher wind ``ram'' pressure and the shock 
shoots out almost radially until the ``ram'' pressure approaches that of the nebula.      
This interpretation suggests that for high $\sigma_0$, the shape of the equatorial 
part of the termination shock is independent on that in highly magnetized polar 
section.

\begin{figure}
\includegraphics[width=\columnwidth]{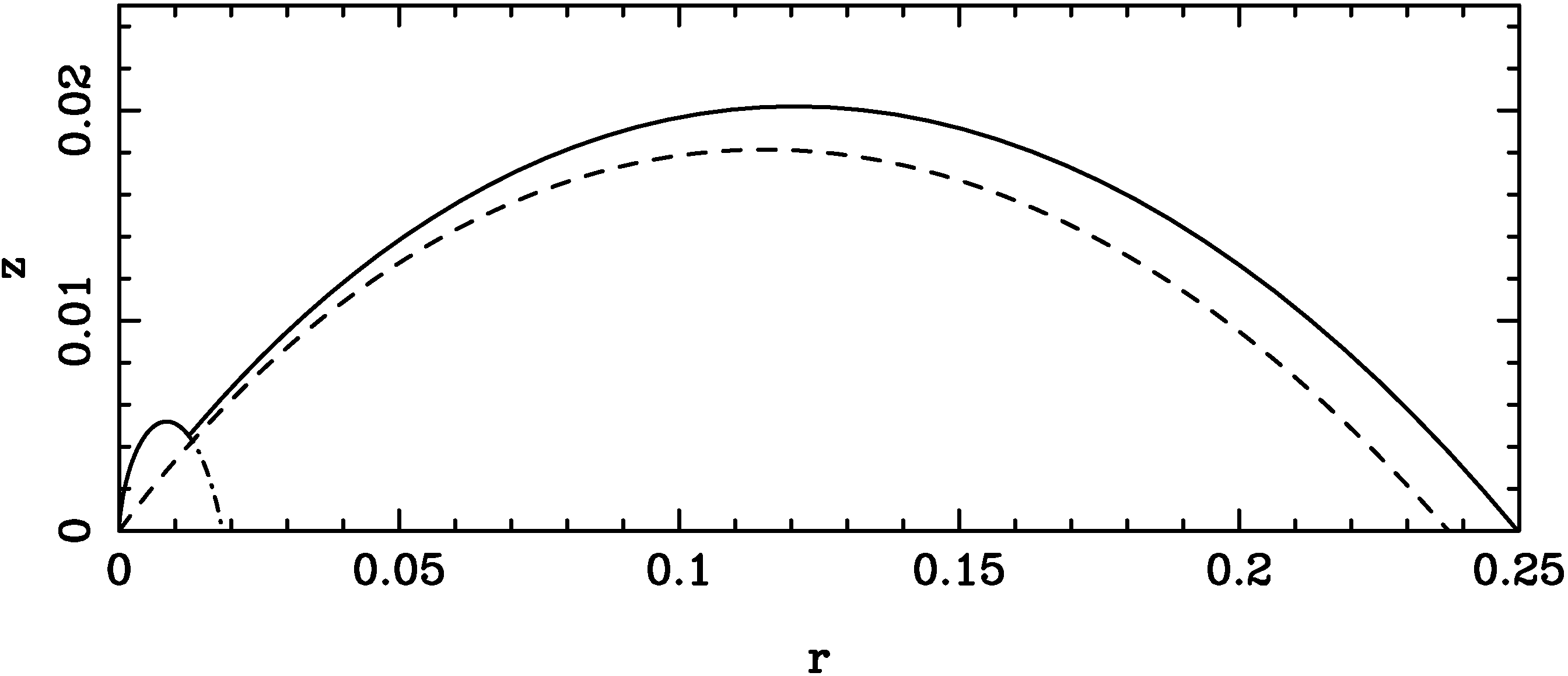}
\caption{Shape of the termination shock for $\sigma_0=10^3$ and $f(\theta)=\sin^2\theta$ 
and $\alpha = 20^o$. The solid line shows the original solution and the dashed one the 
approximate one for the truncated wind and the dash-dotted one the solution corresponding
to $\sigma_1=\sigma_0$.} 
\label{fig:n2app}
\end{figure}

The low ram pressure of the termination shock in the high-$\sigma$ polar region 
and the rapid drop of $\sigma_1$ around $\theta_m$ suggest that, as far as the 
equatorial part of the shock is concerned, one can ignore the presence of the polar 
section of the wind altogether. In this approximation,
the appropriate boundary condition for Eq. (\ref{shock-shape}) is 
\begin{equation}
X(\theta_m)=0\,.
\end{equation}
Figure~\ref{fig:n2app} compares this approximate solution with the original one for 
$\alpha=20^o$, the case in Figure~\ref{fig:all-shapes} with the largest 
unstriped sector. In this case, the difference 
between the solution is expected to be most profound. Yet, as one can see in this figure, 
it is still rather small. This result is particularly welcome as one expects to 
see significant deviation from the uniform pressure distribution of the shocked plasma 
in the polar region where the high-sigma post-shock flow remains supersonic. The exact 
details of the flow in this region should not matter much.

Strictly speaking, our analysis shows that there is no well defined unique shape of the 
termination shock which can be used to predict the emission properperties of 
Crab's inner knot. On the other hand, the dependence on the wind parameters is not 
that strong. With the exception of very small magnetic inclination angle, the shock 
shape is approximately the same as found for the weakly-magnetized wind by 
\citet{2002MNRAS.329L..34L}. For this reason, we will use this shape for the rest 
of our paper.  After small additional rescaling, the shock shape in 
this case is described by
\begin{equation}
    X'^2+X^2 = f(\theta) \,.
\label{shock-eq}
\end{equation}

\begin{figure}
\centering
\includegraphics[width=0.6\linewidth]{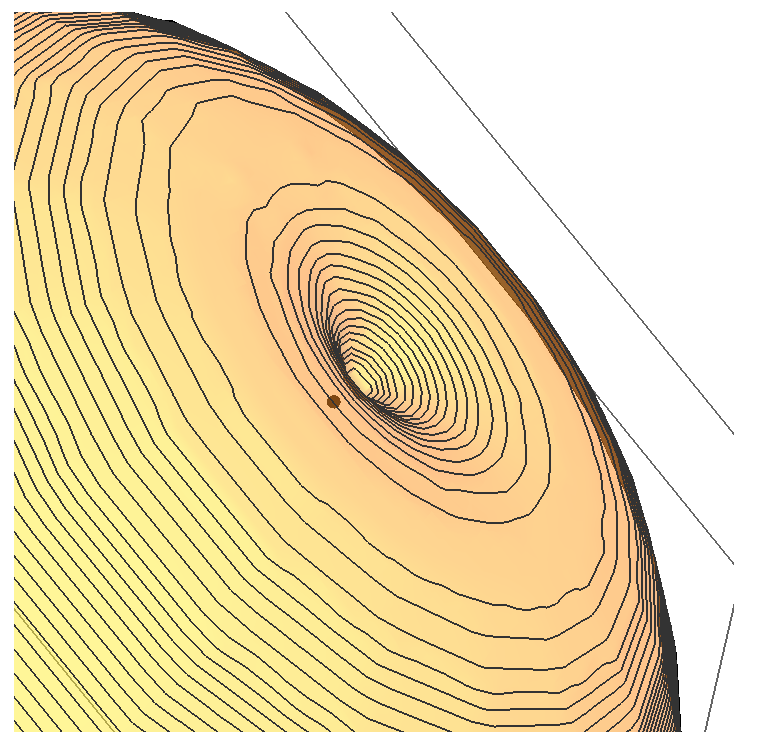}
\caption{View of the polar region of the termination shock for $f(\theta)=\sin^2\theta$ and 
the viewing angle $\theta_{ob}=60^\circ$; the pulsar position is shown by the dot.}
\label{point}
\end{figure}

With $f(\theta)=\sin^n\theta$, the asymptotic solutions of Eq. (\ref{shock-eq}) are
\be 
X\simeq \frac{\theta^2}{2} \left( 1-\frac{7}{48}\theta^2  \right)\, ,
\label{asympt-sol}
\ee
for  $n=2$ and 
\be 
  X \simeq \frac{\theta^3}{3} \left( 1-\frac{7}{90}\theta^2  \right)\, ,
\label{asympt-sol-n4}
\ee
for $n=4$. The corresponding angles of attack are  
\be
   \delta_1 \simeq \theta/2 
\label{delta-app}
\ee  
and 
\be
   \delta_1 \simeq \theta/3
\label{delta-app-n4}
\ee  
respectively. Figure~\ref{point} illustrates how the termination shock with $n=2$ appears 
to a distant observer for the viewing angle of 60 degrees to the symmetry axis.  

\section{Estimates of basic parameters.}
\label{EKS}

As the shocked plasma expands and slows down, its observed emissivity drops. 
Figure~\ref{porth}, based the results of 3D RMHD simulations by \citet{2014MNRAS.438..278P},
illustrates this behaviour. One can see a relatively thin layer of enhanced emissivity 
just above the shock surface. Its thickness is approximately one third of its distance 
from the line connecting the origin (pulsar) and the observer. The main reason for the 
drop of the emissivity with the dinstance from the shock is the reduction of the Doppler 
beaming.           

\cite{2011MNRAS.414.2017K} estimated some of the the knot parameters in the shock 
model, assuming that they are determined by the Doppler boosting of the emission from 
the shocked plasma. In their calculations, they assumed that the velocity of the plasma is 
parallel to the shock surface. Here we do a more careful and extended analysis.  

Assuming a small size of the knot, we first ignore variations of the proper 
emissivity across the knot. In this case, the observed synchrotron emissivity
is \citep[\eg][see also  \S \ref{polariz}]{2003ApJ...597..998L}
\be
\epsilon_\nu \propto {\cal D}^{2+(p-1)/2} |B'_\perp|^{(p+1)/2}\,,
\label{synch-emiss}
\ee
where $p$ is the spectral index of the electron energy spectrum,
$B'_\perp$ is the normal to the line of sight component of magnetic field
in the fluid frame and  ${\cal D}=\Gamma^{-1}(1-v \cos\alpha)^{-1}$
is the Doppler factor. Even if the magnetic field strength is constant over the 
knot, $B'_\perp$ may still vary significantly across the knot due to the relativistic 
aberration of light. However, along the symmetry axis in the plane of the sky 
$B'_\perp=B'$, and it is only the Doppler factor that matters. 

\subsection{Low $\sigma$ at the knot location}

Based on Equation\ref{synch-emiss} one can immediately rule out $\sigma_1\ge 1$ for 
the termination shock at the location of inner knot. The key observational data
here is the clear separation of the knot from the pulsar \citep{2015arXiv150404613R}.
This shows that that the beaming angle $\alpha_d$ is smaller compared to the deflection
angle $\Delta\delta$ of streamlines at the shock. Defining $\alpha_d$ as the angle 
at which ${\cal D}^{2+(p-1)/2}$ reduces the factor of two, we find that for 
the observed spectral index $p\approx2.5$ 
$$
   \alpha_d \approx \frac{1}{2\Gamma_2}. 
$$  
Using the maximum value for the deflection angle and $\Gamma_2$ for high $\sigma_1$
(Equations \ref{gam-high} and \ref{def-angle}) we find that 
$$
   \frac{\alpha_d}{\Delta\delta} \approx 2\sqrt{\sigma_1}\sin\delta_1 = \sqrt{2\sigma_1}\,,
$$ 
where we used $\delta_1=\pi/4$ as the angle of attack with maximal deflection. 
For the case of $\delta_1\ll 1$, we find that 
$$
   \frac{\alpha_d}{\Delta\delta} \approx  \sqrt{\sigma_1}\,. 
$$ 
Both results show that for $\sigma_1\ge 1$ one has ${\alpha_d}>{\Delta\delta}$ and hence 
the pulsar has to be embedded into the knot, in contradiction with the observations. 
Figure~\ref{ratio} shows the ratio of angles as a function of $\sigma_1$ for $\delta_1=\pi/4$ 
and $\pi/20$. One can see, that the dependence of $\delta_1$ is rather week. 
Using \ref{gam-high} and \ref{def-angle} one can show that for $\sigma_1\ll1$
$$
   \frac{\alpha_d}{\Delta\delta} \approx 0.7\,,  
$$
and thus the knot size is comparable with the separation from the pulsar. 
This conclusion does not depend on the shape of the termination shock and thus very robust.

Since $\sigma_1$ is expected to be low only in the striped-wind zone, this allows us to 
conclude that the magnetic inclination angle $\alpha>90^o-\theta_{ob}\approx 30^o$, 
where $\theta_{ob}\approx 60^o$ the observed angle between the line of sight and the rotational axis 
of the Crab pulsar \citep{2004ApJ...601..479N}.     
Based on this result we focus in the rest of the paper on the case of low $\sigma_1$.

\subsection{Separation from the pulsar}

\begin{figure}
\centering
\includegraphics[width=\columnwidth]{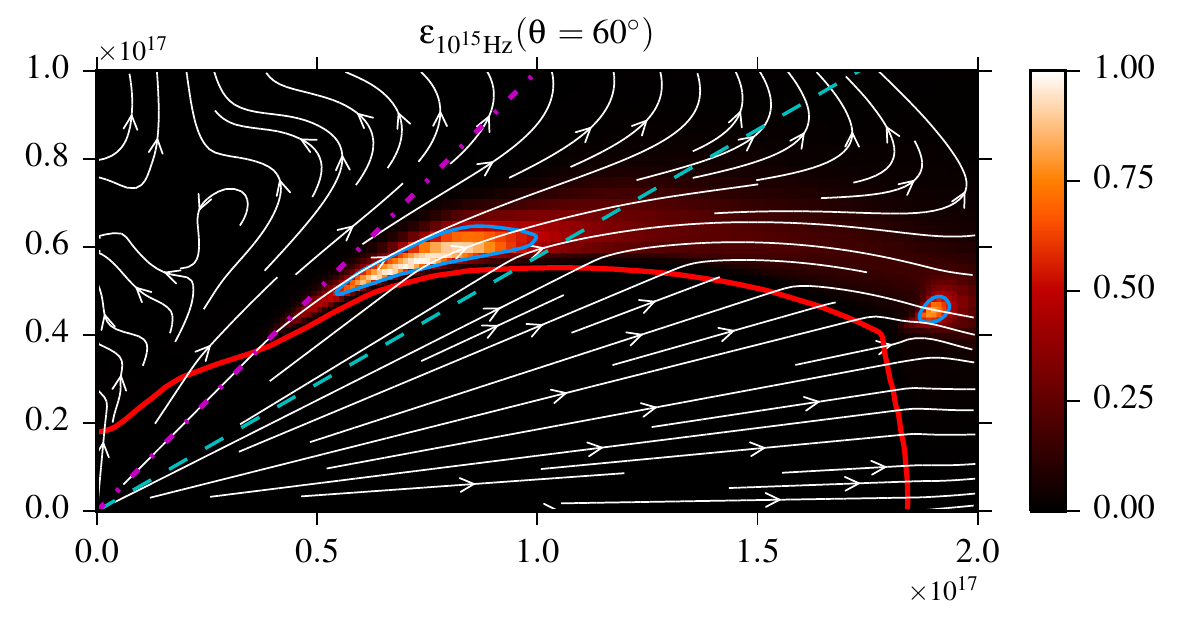}
\caption{The observed optical emissivity of the post-shock flow in the 3D RMHD 
simulations of \citet{2014MNRAS.438..278P}. The arrowed lines are the 
instantaneous stream lines. The dashed line is the line of view and the blue 
curves show the regions of enhanced observed emissivity.  
}
\label{porth}
\end{figure}

\begin{figure}
\centering
\includegraphics[width=0.8\columnwidth]{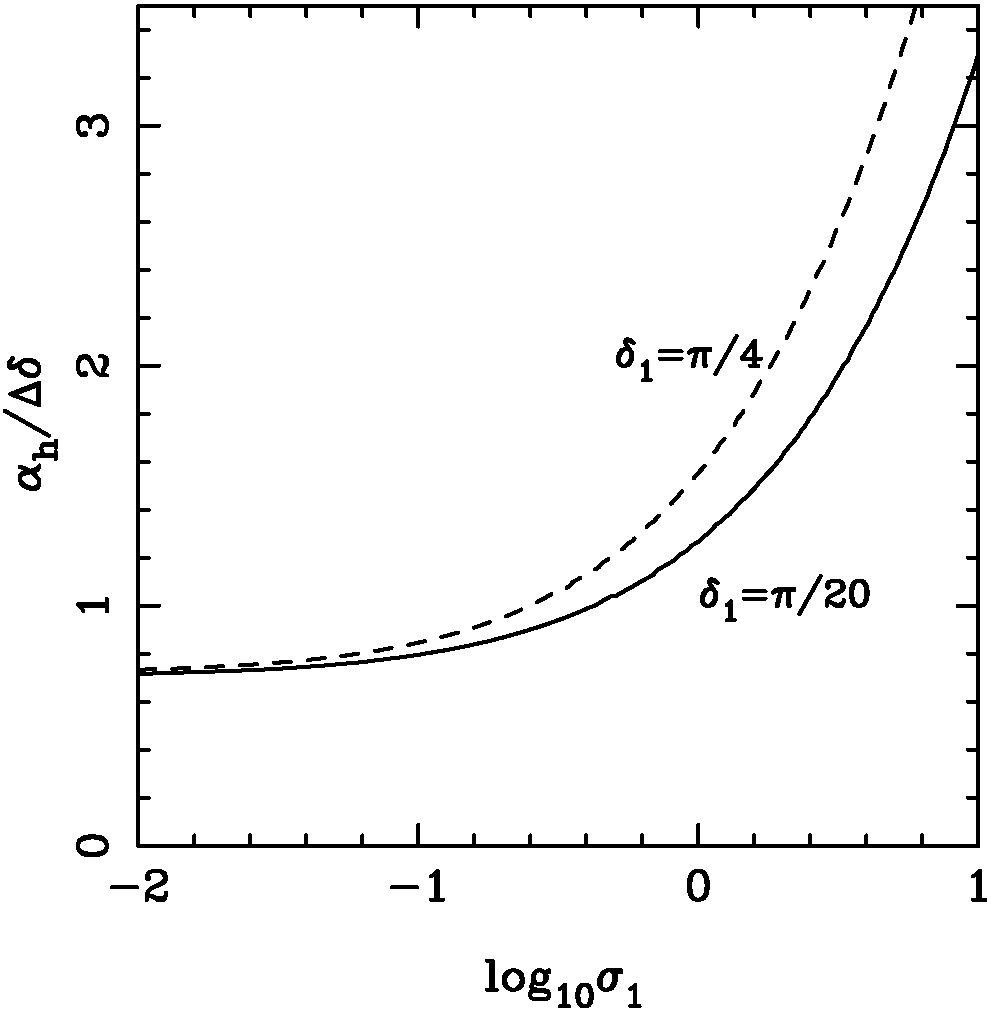}
\caption{The ratio of the beaming and deflection angles. 
}
\label{ratio}
\end{figure}

\begin{figure*}
\centering
\includegraphics[width=0.7\linewidth]{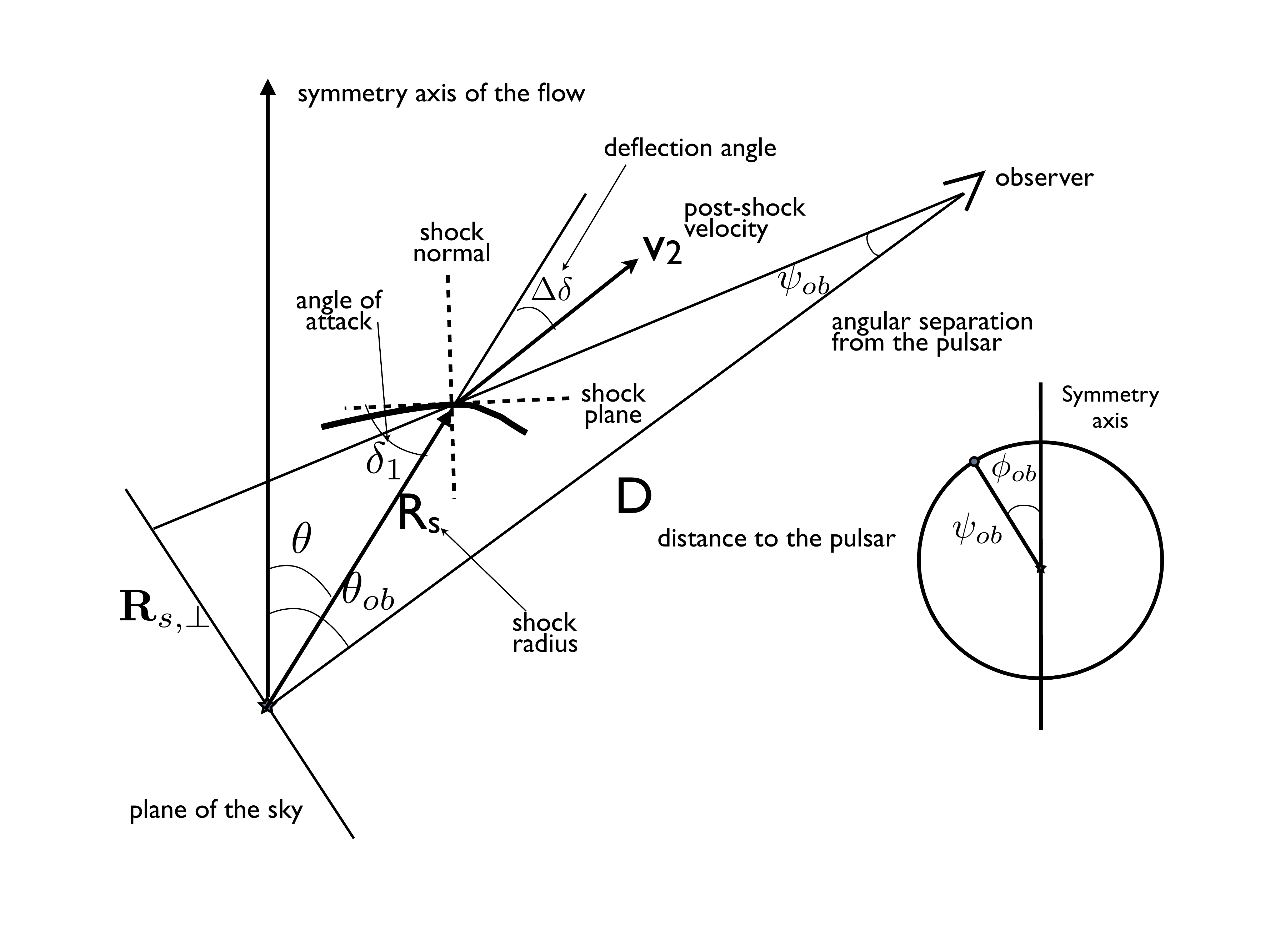}
\caption{Geometry of the shock in the plane defined by the location of an observer, 
the pulsar and a point on the shock surface.}
\label{geom1}
\end{figure*}

The brightness peak of the knot corresponds to the point where the deflected 
streamline points directly towards the observer. The polar angle of this point 
$\theta_k=\theta_{ob}-\Delta\delta$ (see Figure~\ref{geom1}).
Using Eq.~(\ref{def-angle})
in the limit of small angles and Eq.~(\ref{delta-app}), we find that

\be
    \Delta\delta \approx (1-\chi)\delta_1 \approx (1-\chi)\frac{\theta}{2} 
\label{Dd}
\ee
and 

\be
   \theta_{k}=\frac{2}{3-\chi}\theta_{ob} \,. 
\ee
The angular separation between this point and the pulsar in the plane of the sky 
is 

\be
   \psi_k \approx \frac{R_k}{D} \Delta\delta \, ,
\ee
where $D$ is the distance to the pulsar.
Denoting as $R_{ts}$ and $\psi_{ts}$ the linear equatorial radius of the 
termination shock and its angular size in the plane of the sky respectively, 
\be 
   \frac{\psi_k}{\psi_{ts}} = \frac{R_k}{R_{ts}}\Delta\delta \,.
\ee

For $\sigma\ll1$, one has $\chi\approx 1/3$ and hence 
\be
   \Delta\delta \approx \frac{1}{4}\theta_{ob}\,, \quad  
\theta_k=\frac{3}{4} \theta_{ob}\,.
\ee
Now $R_k=0.28$ and $\psi_{ts} = 11.4 \psi_{p} \approx 7.4\arcsec$.
This is approximately equal to the radius of the Crab's halo \citep{1995ApJ...448..240H} 
and almost twice as small compared to the radius of its X-ray ring.
For the shock shape function $f(\theta)=\sin^4\theta$, one obtains 
$R_k=R_s(\theta_k) = 0.14 R_0$,  $R_{ts} =R_s(\pi/2) =0.70 R_0$ and 
$\psi_{ts} = 20\psi_{p} \approx 13\arcsec$. Thus, given the uncertainty of the shock shape, the 
theory and observations are quite consistent in the limit of low $\sigma_1$.

\begin{figure*}
\includegraphics[width=0.3\linewidth]{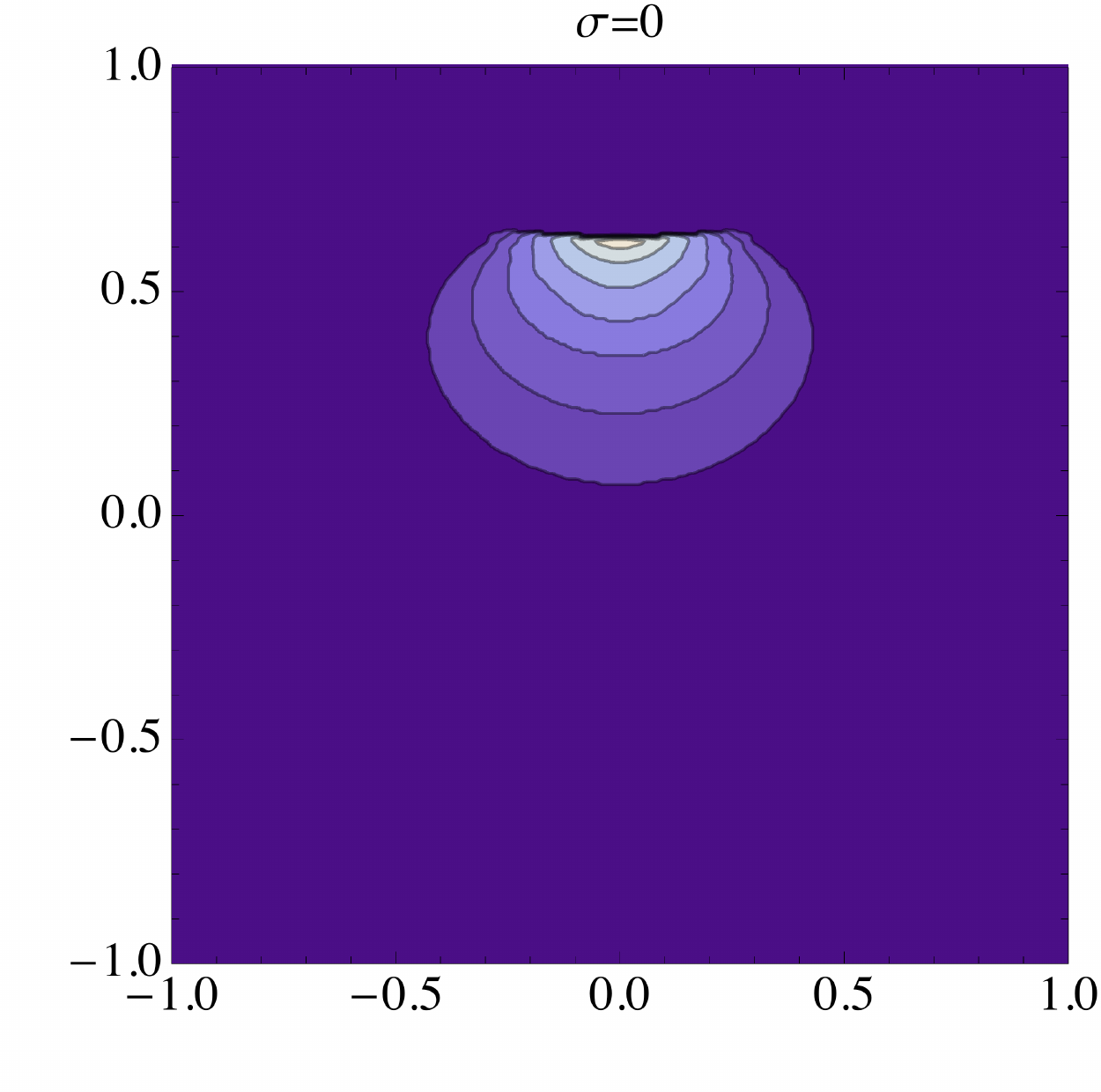}
\includegraphics[width=0.3\linewidth]{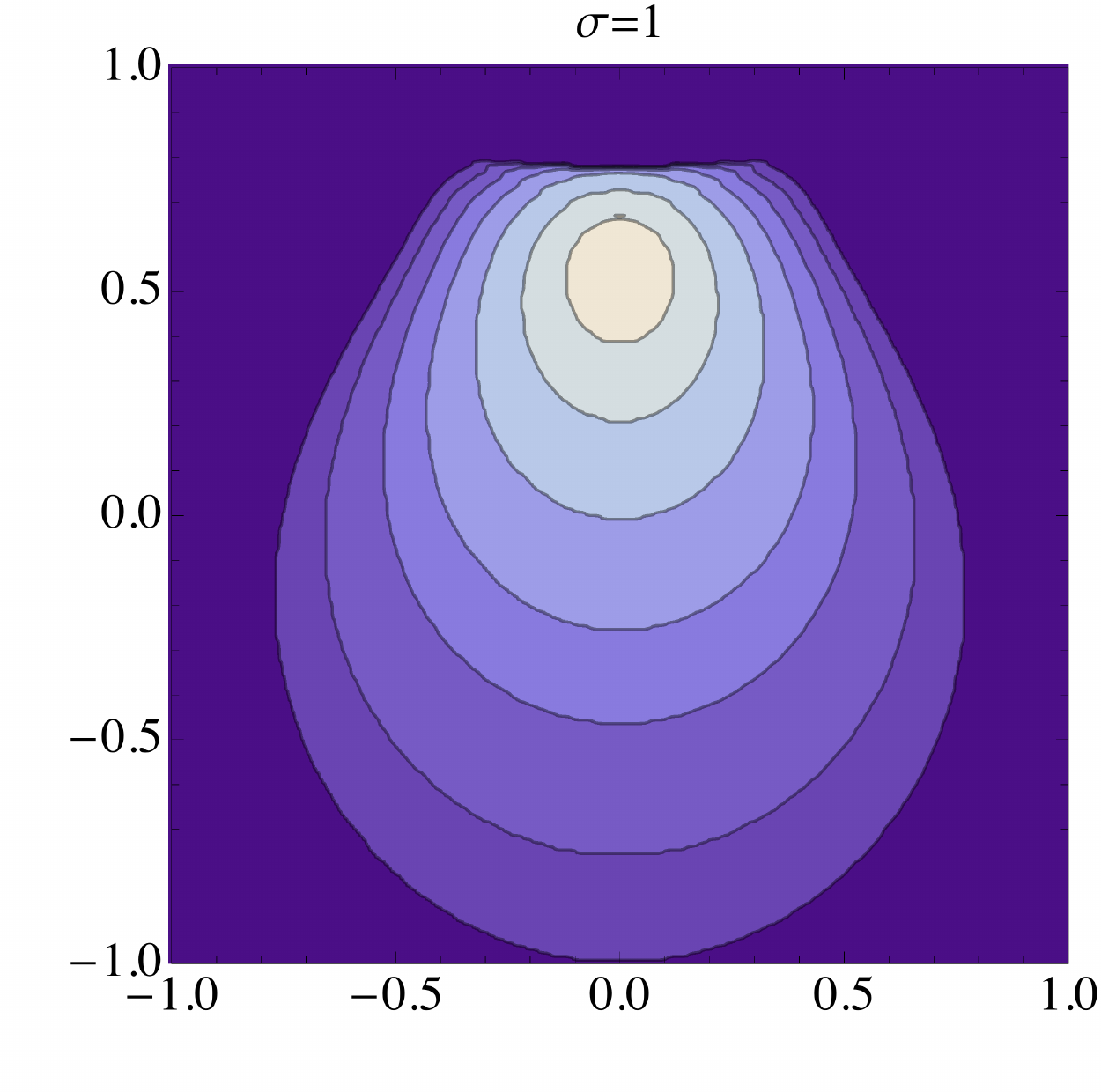}
\includegraphics[width=0.3\linewidth]{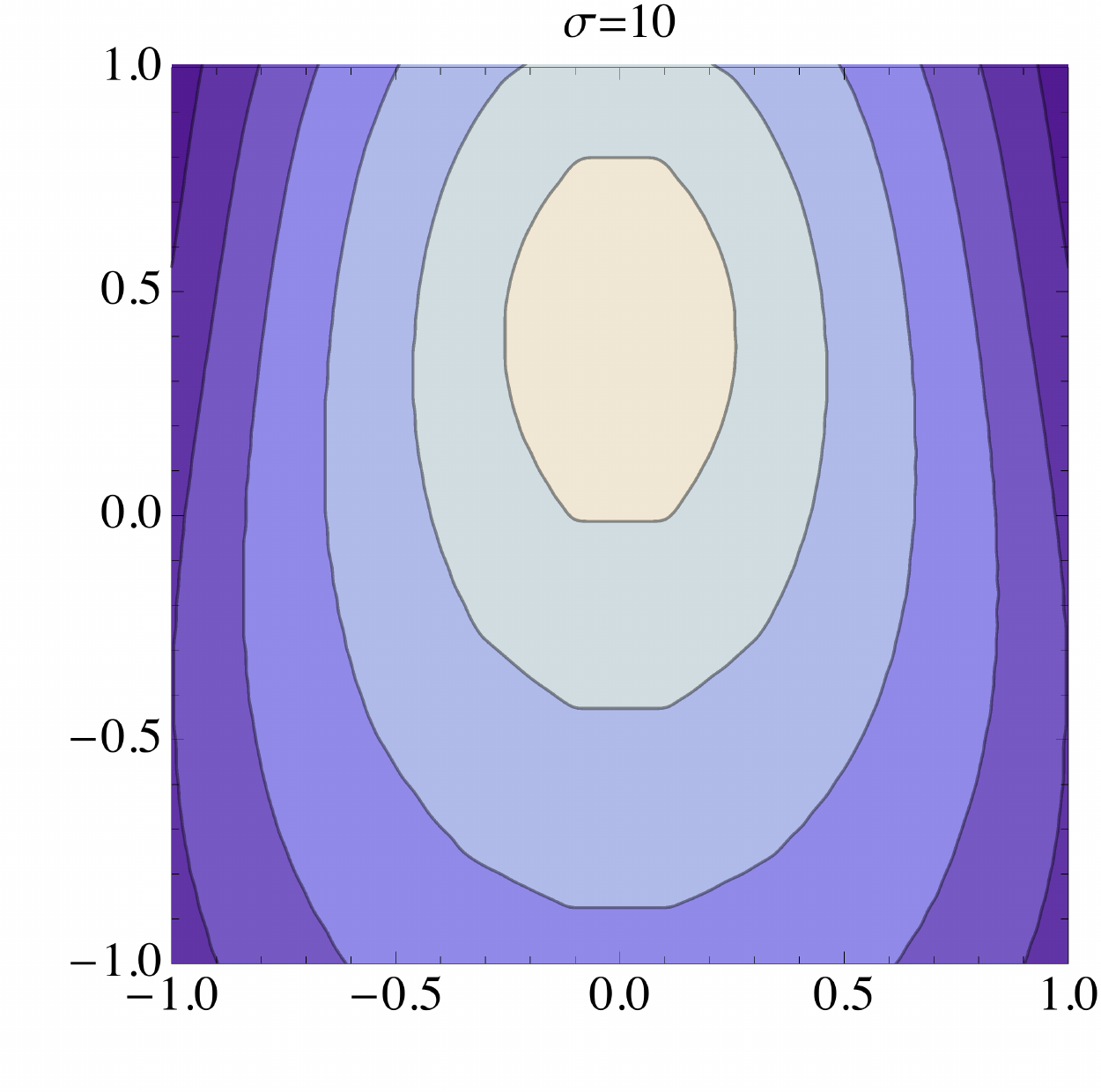}
\includegraphics[width=0.05\linewidth]{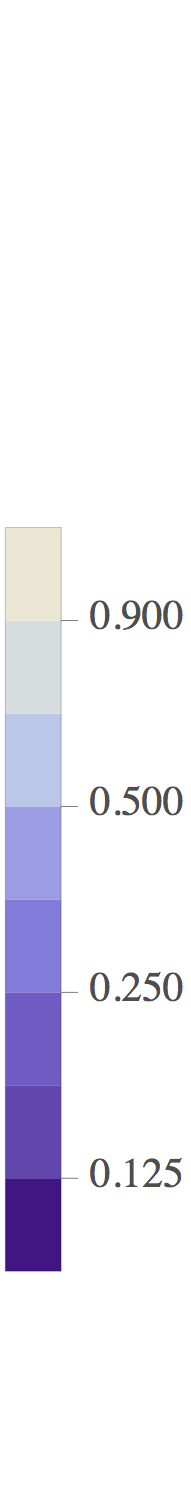}
\caption{Synthetic emission maps of the knot for $\sigma_1=0,1,10$ (from left to right) 
and $f(\theta)=\sin^2\theta$. In all cases, the distances are 
in arcseconds, the pulsar is located at the
origin and the emission peak is at $0.7\arcsec$ from the pulsar.  
 The peak emissivity is normalized to unity and the third contour corresponds to 
one half of the peak value.  
For high $\sigma\geq 1$ the inner knot is both very broad and elongated in radial direction, 
which is inconsistent with the observations. The contours start from 0.9 of the peak value 
and decrease by the factor of $\sqrt{2}$ thereafter.  }
\label{mapSigma}
\end{figure*}

\begin{figure*}
\includegraphics[width=0.3\linewidth]{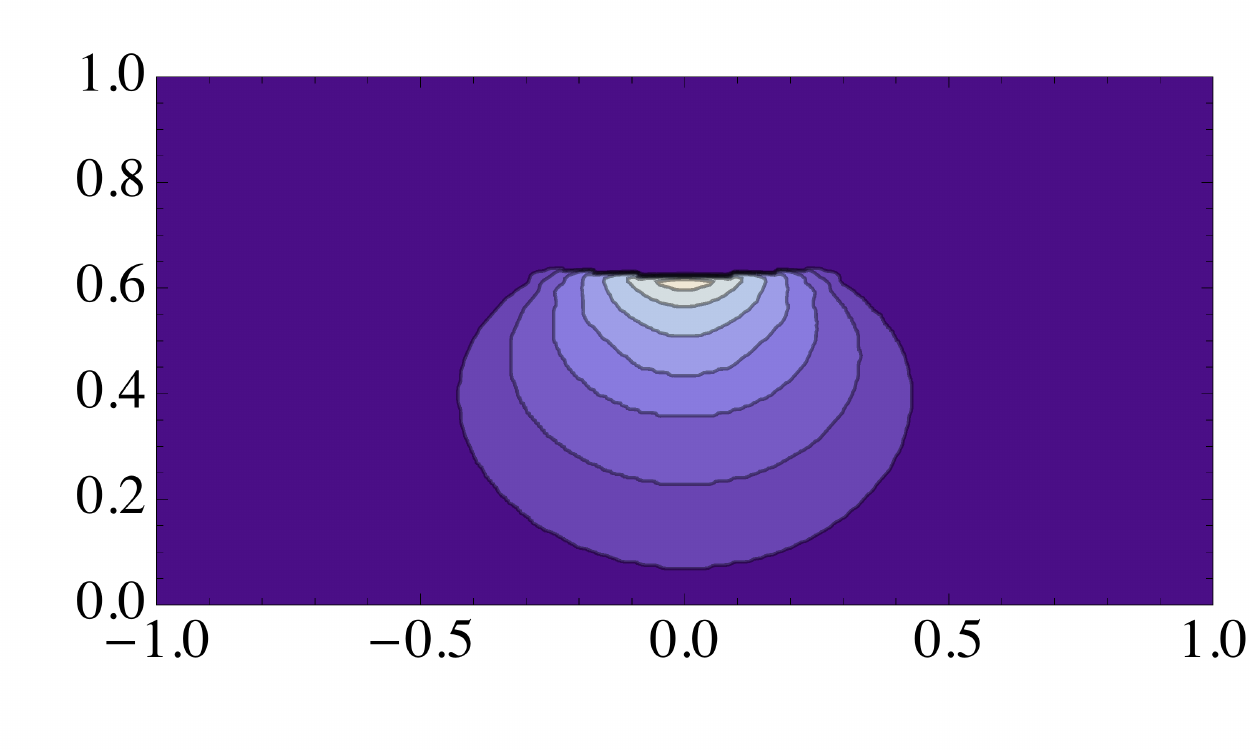}
\includegraphics[width=0.3\linewidth]{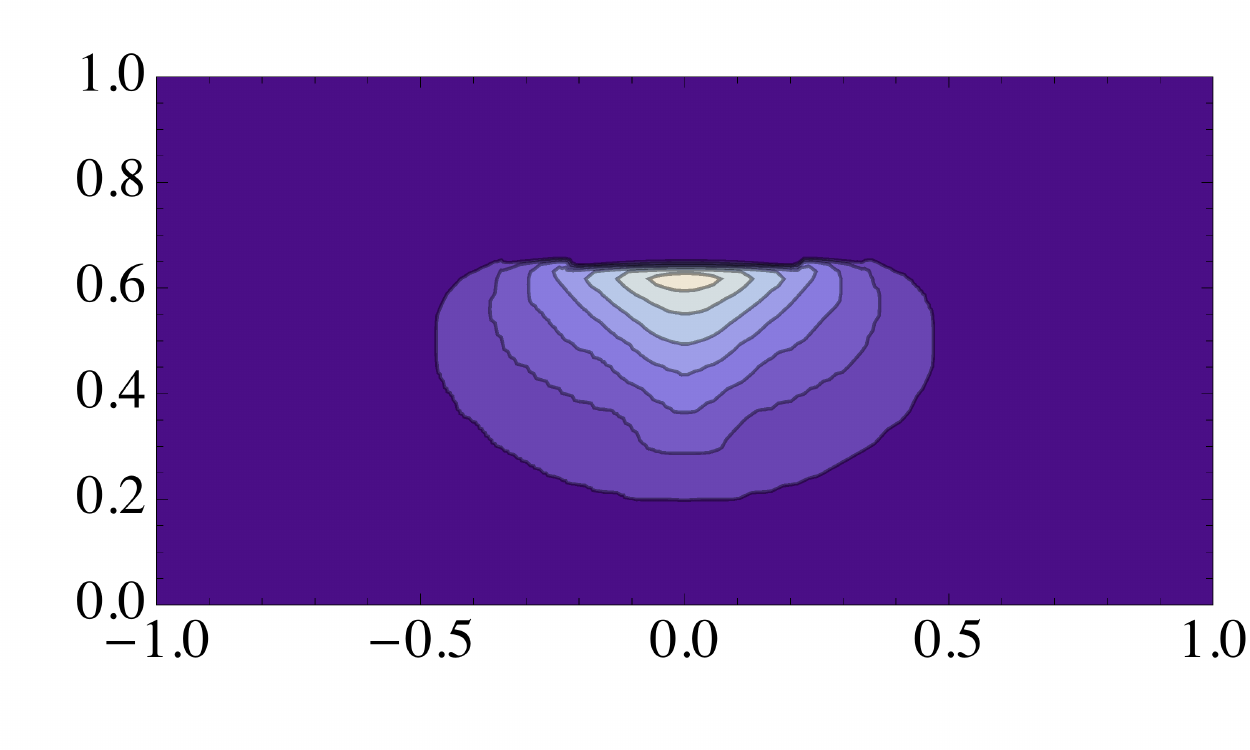}
\includegraphics[width=0.3\linewidth]{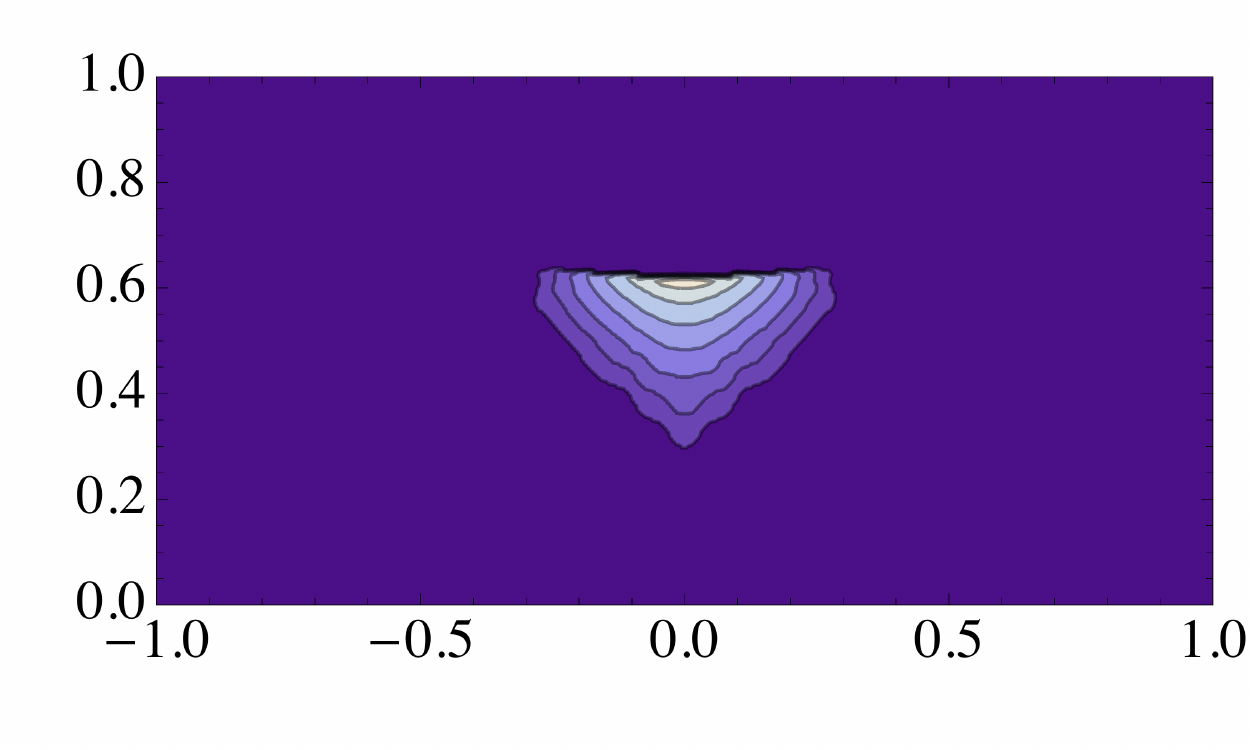}
\includegraphics[width=0.05\linewidth]{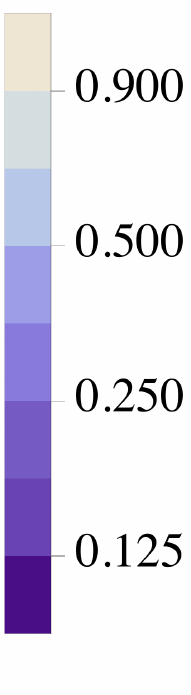}
\caption{Sensitivity of the synthetic images to the shock model. 
{\it Left panel}:  shock shape $f(\theta)=\sin^2\theta$ and uniform proper emissivity; 
{\it Center panel}: shock shape $f(\theta)=\sin^4\theta$ and uniform proper emissivity; 
{\it Right panel}: shock shape $f(\theta)=\sin^2\theta$ and proper emissivity scaling 
with the spherical radius as $\propto R^{-2}$. In all the cases $\sigma_1=0$.
The contours start from 0.9 of the peak value 
and decrease by the factor of $\sqrt{2}$ thereafter.
}
\label{mapSigma1}
\end{figure*}

\subsection{Transverse size}
\label{tr-size}

The full half-brightness transverse size of the knot can be estimated as 
\be
   \Delta\psi_\perp= 2 \alpha_h \frac{R_k}{D}\,,
\ee
where $\alpha_h$ is the angle between the line of sight and the velocity vector at 
the point on the shock, with the same position on the shock-defining curve as 
the center of the knot, where the emissivity is reduced by the factor of two because 
of the Doppler effect and the relativistic aberration of light. 
Thus, we have 
\be
  \frac{\Delta\psi_\perp}{\psi_k} = \frac{2\alpha_h} {\Delta\delta} \,.
\label{nsize}
\ee

The observed synchrotron emissivity is given by Eq. (\ref{synch-emiss}).  
Provided the knot size is small, one can assume that the 
magnetic field is uniform and write $B'_\perp=B'\cos\alpha '$, where the $\alpha'$ is 
the angle between the stream line and the line of sight in the fluid frame. 
The relativistic aberration of light gives
\be    
\cos\alpha' =\frac{\cos\alpha -v}{1-v\cos\alpha}\,. 
\ee 
Substituting this into Eq. (\ref{synch-emiss}), we find that 
\be
\epsilon_\nu \propto (1-v\cos\alpha)^{-(p+2)} (\cos\alpha-v)^{(p+1)/2}\,.
\ee
Approximating $v\simeq 1-1/2\Gamma_2^2$ and $\cos\alpha \simeq 1-\alpha^2/2$, 
this reads 
\be
\epsilon_\nu \propto (1+x^2)^{-(p+2)} (1-x^2)^{(p+1)/2}\,,
\ee
where $x=\Gamma_2\alpha$. For the observed $p=2.6$, this equals to one half for 
$x\approx 0.33$ and. Thus $\alpha_h \approx 0.33/\Gamma_2$ and Eq. (\ref{nsize}) reads
\be
  \frac{\Delta\psi_\perp}{\psi_k} \approx \frac{0.66}{\Gamma_2\Delta\delta} \,.
\label{nsize1}
\ee
Substituting into the last equation the expressions (\ref{gamma2}) and (\ref{def-angle}) 
in the approximation of small $\delta_1$, we finally obtain 
\be  
\frac{\Delta\psi_\perp}{\psi_k} \approx 0.66 \fracp{1+\chi}{1-\chi}^{1/2} \,.
\label{perp-size}
\ee
This is a monotonically increasing function of $\chi$ and has the absolute minimum 
value $(\Delta\psi_\perp/\psi_k)_{min} = 0.9$ reached for $\sigma_1=0$ ($\chi=1/3$). 
For $\sigma_1=10$ this gives $\Delta\psi_\perp/\psi_k =4.3$. Observational measurements of the 
knot parameters are complicated by its small size and proximity to the bright 
Crab pulsar.  Depending of the method used, the transverse size of the knot in HST 
images varies from  $\Delta\psi_\perp \approx 0.3\arcsec$  to  $0.56\arcsec$, 
whereas $\psi_k \approx 0.65\arcsec$ \citep{2015arXiv150404613R}. This rules
out high $\sigma_1$ and favors $\sigma_1 \ll 1$ once more.  The synthetic images of 
the knot presented in Section~\ref{images}, give a somewhat smaller size compared 
to what follows from Eq. (\ref{perp-size}).  

\section{Synthetic images}
\label{images}

\begin{figure}
\begin{center}
\includegraphics[width=0.4\textwidth]{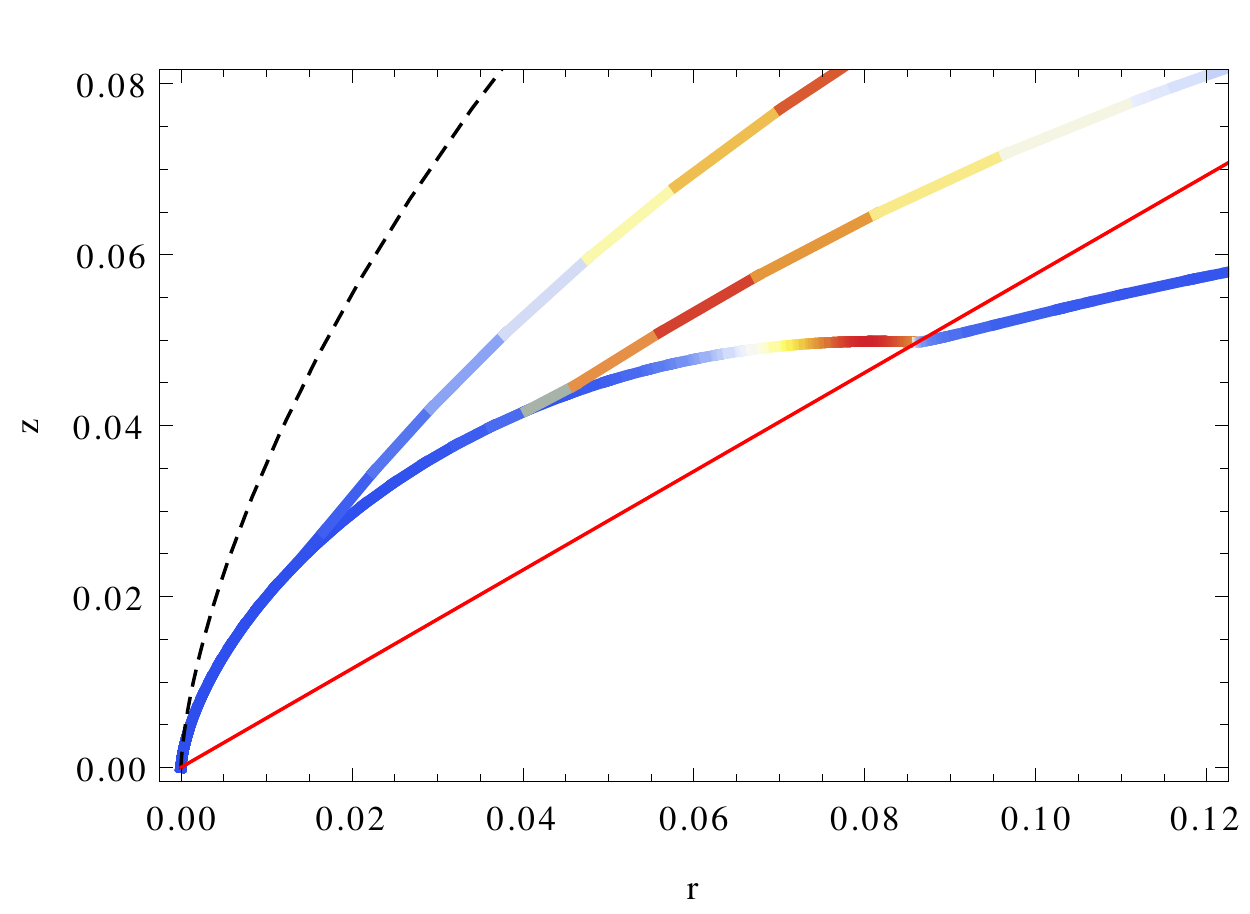}
\caption{Shock contour for obliqueness of $\pi/3,\pi/4,\pi/6$ (top to
  bottom) for wind magnetization of $\sigma_0=10$ and
  $f(\theta)=\sin^2\theta$.  The hydro shock solution is indicated
  dashed black.  The shock-contours are colored with the Doppler
  factor cubed.  Red contour describes the los to the Pulsar for the
  assumed inclination angle $\theta_{\rm ob}=60^\circ$.  Right-hand
  panel is a zoom-in.  One can see the double-humped structure of the
  termination shock, leading to a emission closely aligned with the
  los in the case of $\alpha=\pi/4$.}
\label{fig:varshape-shock}
\end{center}
\end{figure}

\begin{figure*}
\begin{center}
\includegraphics[width=0.3\textwidth]{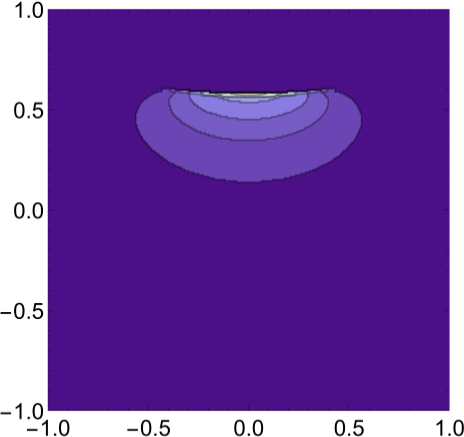}
\includegraphics[width=0.3\textwidth]{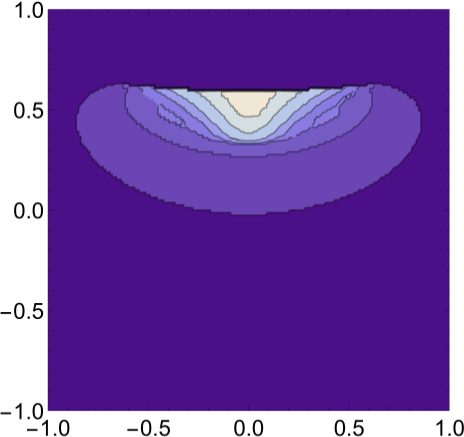}
\includegraphics[width=0.3\textwidth]{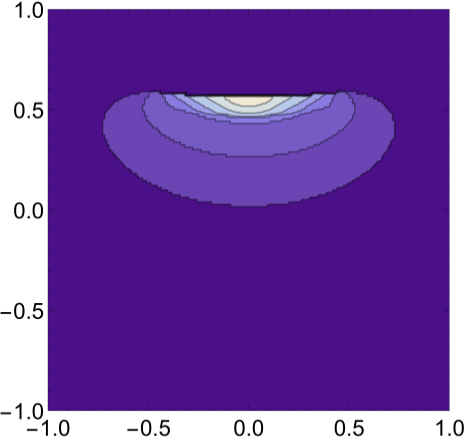}
\includegraphics[width=0.05\textwidth]{figures/colorbar.png}
\includegraphics[width=0.3\textwidth]{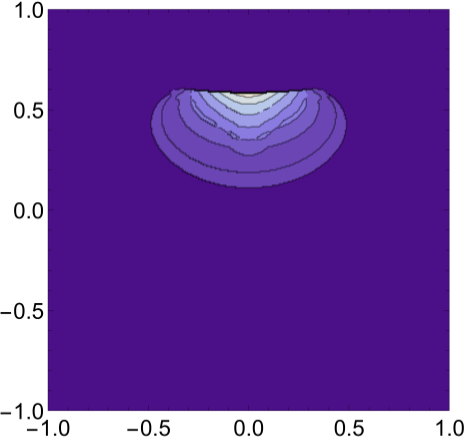}
\includegraphics[width=0.3\textwidth]{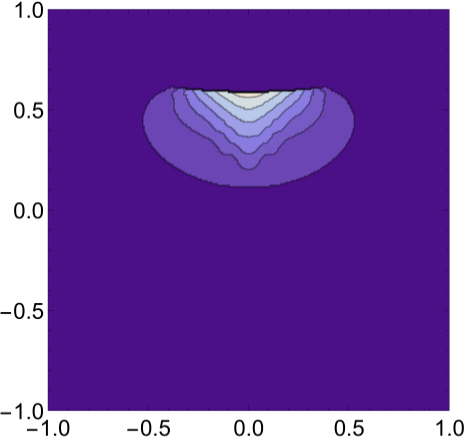}
\includegraphics[width=0.3\textwidth]{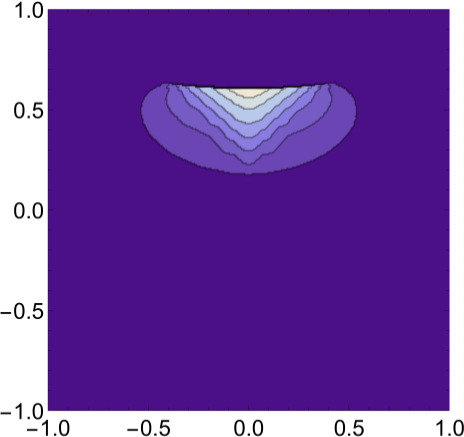}
\includegraphics[width=0.05\textwidth]{figures/colorbar.png}\\
\caption{Emissivity maps for the striped-wind shock shape.  
Top row: $\alpha=\pi/4$. Bottom row: $\alpha=\pi/3$.  
Left column: $\sigma_0=10$, $n=2$;
Middle column: $\sigma_0=100$, $n=2$;
Right column: $\sigma_0=10$, $n=4$.
The contours start from 0.9 of the peak value
and decrease by the factor of $\sqrt{2}$ thereafter.
}
\label{fig:varshape-emiss}
\end{center}
\end{figure*}

In this section we construct two-dimensional ``images of the knot''. Obviously, 
in order to obtain the brightness distribution we need to integrate the emissivity
along the line of sight. However, from the shock geometry we can only conclude how 
it is distributed over the shock surface. Therefore, in this section we start by
constructing images of this surface and later study the effects of finite thickness 
of the emitting layer. We expect a longer geometrical length of the emitting region 
along the line tangent to the shock surface. This factor would  make the knot more 
compact along the symmetry axis in the image. On the other hand, the finite thickness of 
the emitting layer would tend to increase the knot size in this direction.   

For all images presented in the paper we use $\theta_{ob}=\pi/3$.  

\begin{figure*}
\centering
\includegraphics[width=0.49\linewidth]{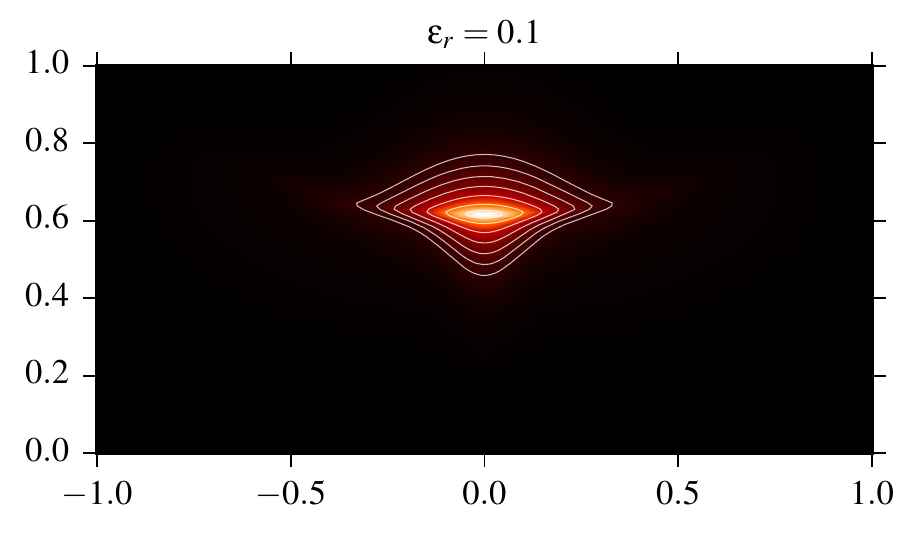}
\includegraphics[width=0.49\linewidth]{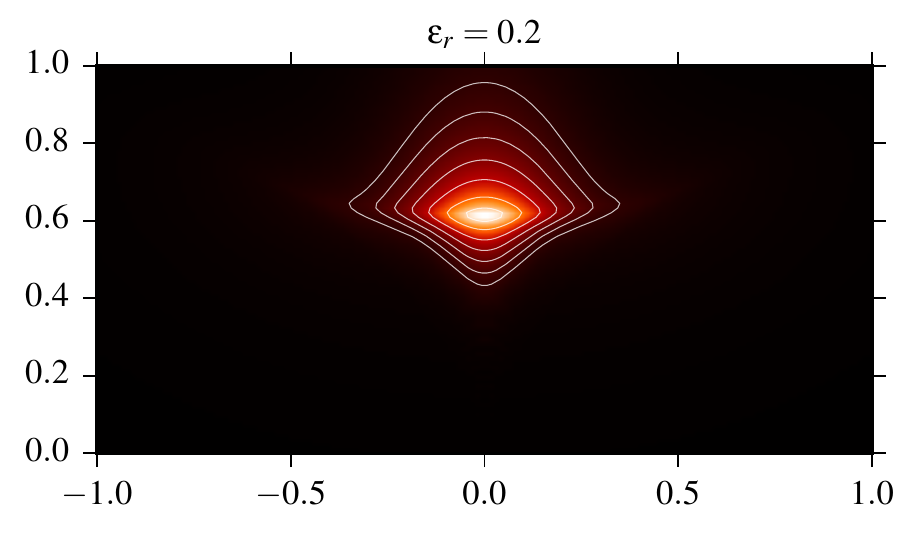}\\
\includegraphics[width=0.49\linewidth]{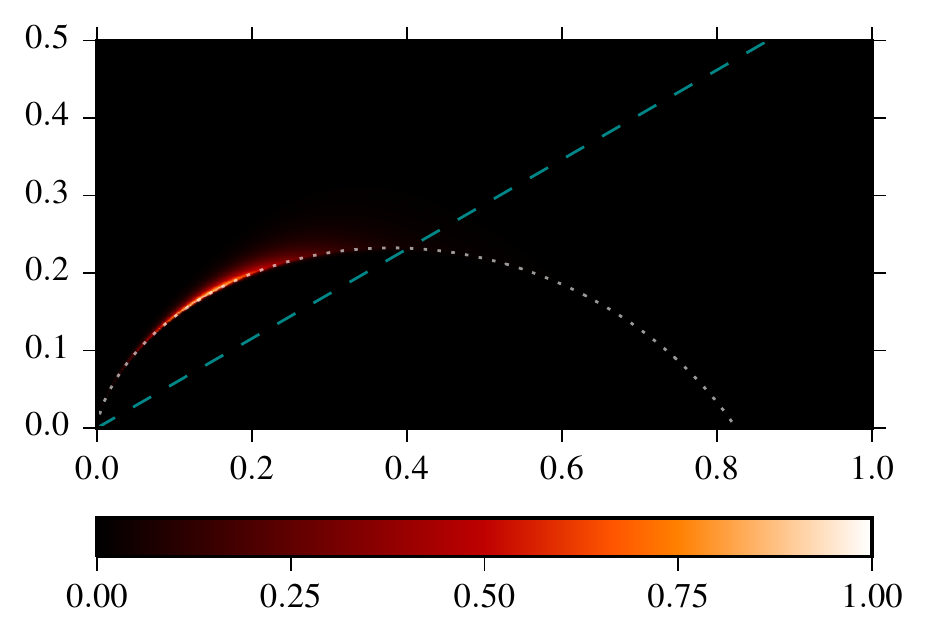}
\includegraphics[width=0.49\linewidth]{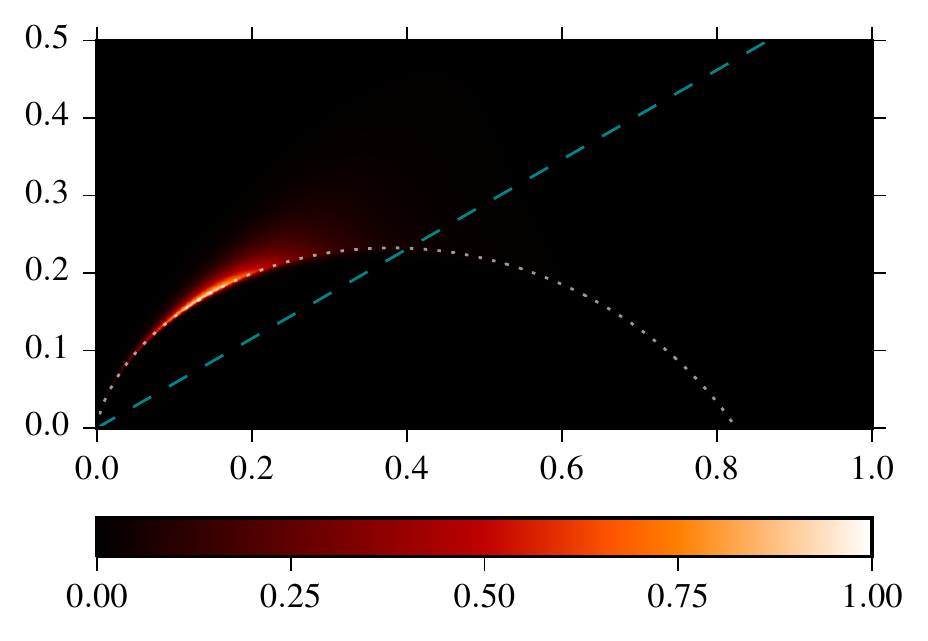}
\caption{Knot images resulting from emitting region of finite thickness
\textit{(top)} and corresponding emissivity distributions \textit{(bottom)}.  
The dashed line shows the line-of-sight and the white dotted line
traces the assumed shape of the termination shock.  
The contours start from 0.9 of the peak brightness
and decrease by the factor of $\sqrt{2}$ thereafter.
}
\label{fig:emiss-thick}
\end{figure*}

\subsection{Emissivity maps}

Given the shock shape and the ``upstream'' magnetization parameter $\sigma_1$  
one can determine the post-shock flow direction and its Lorentz factor, as well 
as the angle $\mu'=\pi/2-\alpha'$ between the line of sight and the magnetic field 
in the fluid frame. These allow us to compute the purely geometrical component
of the synchrotron emissivity over the shock surface. Namely, Eq.\ref{synch-emiss} 
gives us that 
\be
\epsilon_\nu \propto \epsilon_{geom}={\cal D}^{2+(p-1)/2} | \sin \mu'|^{(p+1)/2}\,.
\label{synch-emiss-1}
\ee
Next, we project this distribution of $\epsilon_{geom}$ on the plane of the sky. 
The main contribution to the knot emission comes from the closest to the observer 
section of the shock surface.  Due to the non-spherical shock geometry, the line of 
sight may intersect this section twice. In this case, we sum the contributions from 
both these points. Next we rescale the image so that the maximum is located at 
$0.7$ arcsec from the pulsar.  

As an illustration, Figure~\ref{mapSigma} shows the results for the shock shape 
described by Eq.\ref{shock-eq} with $n=2$ and constant magnetization parameter 
$\sigma_1 =0,1,10$.  
One can see that only in the case $\sigma_1=0$ thereis a clear separation 
between the knot and the pulsar, in full agreement with the results of 
Sec.\ref{EKS}. The plots also confirm the conclusion of 
Sec.\ref{EKS}, that high $\sigma_1$ models result in radially elongated 
elongated images, which is in conflict with the observations. 

In Figure~\ref{mapSigma1}, we compare the results for $n=2$ and $n=4$. One 
can see that the difference between the model is not dramatic -- in the $n=4$ model, 
the knot is a little  bit more tangentially elongated. The proper shock 
emissivity may reduce with the distance from the pulsar reflecting the reduced 
wind power. To probe the inportance of this factor, we also considered the model 
where the shock emissivity scales as $R^{-2}\epsilon_{geom}$ -- the 
results are shown in the right 
panel of Figure~\ref{mapSigma1}. One can see that the knot becomes significantly more 
compact and less elongated in the radial direction. This image is closer to those of the 
Crab's inner knot, which is approximately 2:1 in size
(tangential over radial), while its separation from the pulsar, $\sim
.65$ arcsec is much larger than it's radial width ($0.15$ arcsec in
the HST image and somewhat larger, $0.35$ arcsec in the Keck image),
\citep{2015arXiv150404613R}.

Finally, we have also explored the case of the shock shape of striped wind 
described by Eq.~\ref{shock-shape-eq}, with $\sigma_1$ varying according to 
Eq.\ref{eff-sigma}. In all models described here, the shock emissivity is 
$\epsilon = R^{-2}\epsilon_{geom}$. 

For large magnetic inclination angle $\alpha>\theta_{ob}$, the shock shape  
is very similar to the our ``standard'' one. Moreover, $\sigma_1\ll 1$ and 
hence the images are not much different from those shown in Figure~\ref{mapSigma1}.   
For small magnetic inclination angle $\alpha<\theta_{ob}$ the knot emission comes 
from the inner lobe of the shock, where the shock shape is exactly the same as the 
standard one.  However, $\sigma_1$ is very high now, leading to images which are in 
stark conflict with the observations (like the one in the left panel of 
Fig.~\ref{mapSigma}).  

The most interesting is the case with $\alpha\approx\theta_{ob}$, where the 
knot emission comes from the transitional section of the shock where $\sigma_1$
varies rapidly and the shock surface is closely aligned with the line of sight 
(see Figure~\ref{fig:varshape-shock}). In Figure~\ref{fig:varshape-emiss} we show
the results for $\alpha=\pi/3$ and $\pi/4$. One can see that for $\alpha=\pi/3$, 
a the images can show a very high degree of elongation in the transverse direction. 
However, for $\alpha=\pi/4$, this elongation is no longer seen. 
The results confirm our expectation that the differences between standard 
shock shape (Eq.\ref{shock-shape}) and that of the striped wind 
(\ref{shock-eq}) are not dramatic and the conclusions based on the models 
with standard shape are quite robust. 

\cite{2015arXiv150404613R} also pointed out that the Crab's knot
could be a bit convex away from the pulsar (the ``smily face''). In 
the synthetic synchrotron maps presented in this section, the distant side of 
the knot has a sharp edge slightly convex the other way. This feature 
reflects the curvature of the folding edge of the shock surface 
projection onto the plane of the sky.  However, in the images based on 
the RMHD numerical simulations of PWN the edge curvature is washed away 
\citep{2014MNRAS.438..278P}.  In these simulations, the emission comes from 
a layer of finite thickness downstream of the termination shock 
(see Figure~\ref{porth}). As we show next, this can be an important factor in 
determining the detailed shape of the knot.

\subsection{Brightness distribution}
\label{layer}

In order to probe the effect of finite thickness of the emitting layer,
we need a model of volume emissivity away from the shock surface. 
Our starting point is the emissivity on the shock surface, which we assume to be 

\be
\epsilon_0(R,\theta,\phi) 
\propto R^{-2}{\cal D}^{2+(p-1)/2} |\sin \mu'|^{(p+1)/2}. 
\label{eq:emiss-epsilon}
\ee
The emissivity outside of the surface is then modeled as 
\be     
  \epsilon(R,\theta)=\epsilon_0(R,\theta,\phi)\delta_{\epsilon} (R,\theta)\,,
\label{emiss-volume}
\ee 
where the function $\delta_{\epsilon} (R,\theta)$ provides spreading about the 
surface. We choose it to be  
\be
\delta_{\epsilon}(R,\theta)= \exp(- |R-R_s(\theta)|/ (\epsilon_r R_s(\theta)))
\ee
for $R>R_s(\theta)$ and 
\be 
\delta_{\epsilon}(R,\theta)  \tanh(-4|R-R_S(\theta)|/(\epsilon_r R_s(\theta))) +1 
\ee
for  $ R\le R_s(\theta)$, where $R=R_s(\theta)$ is the shock radius. 
The parameter $\epsilon_r$ controles the relative thickness parameter. 
The factor of $4$ in the argument of $\tanh$ provides much faster drop of 
emissivity in the direction towards the pulsar. 

In Figure \ref{fig:emiss-thick} illustrates the results obtained for the 
shock shape parameter $n=2$ with $\sigma_1=0$. As one can see, 
the ``frown'' turns into a ``smile'' already for $\epsilon_r=0.1$.  
At this point, the emissivity is still a sharp layer attached to the shock.  
Notwithstanding the ad-hoc nature of this simple model, the results
indicate that the shape of the knot is very sensitive to the
downstream flow.  We hence suggest that a modeling of the knot's
shape must take into account the flow in the post-shock layer.

\section{Polarization}
\label{polariz}

Given the velocity field and the assumed magnetic structure of
the flow (azimuthal) we can also calculate the polarization
signature. In order to do this properly, the relativistic aberration of light
has to be taken into account \cite{2003ApJ...597..998L}. In our case, 
the calculations are slightly different due to the different geometry 
of the problem. The details can be found in Appendix~\ref{app-emission}.

\begin{figure}
\centering
\includegraphics[width=\linewidth]{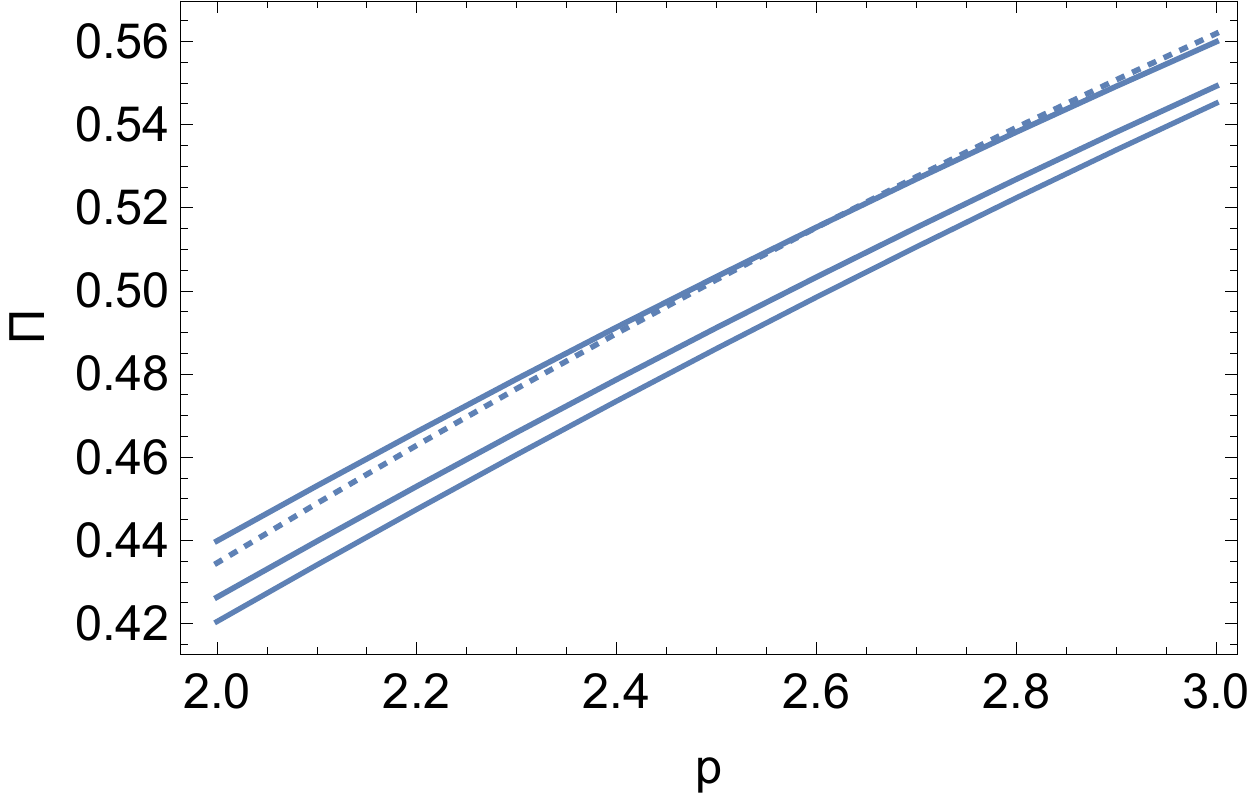}
\caption{Polarization of the shock integral emission in the case of zero thickness 
as a function of the particle spectral index $p$.   
Solid curves correspond to the integration over the whole shock surface for 
models with $\sigma_1= 1,0.1,0$ (top to bottom) and constant rest-frame emissivity.  
The dashed line shows the result for spherical surface obtained 
in \citet{2003ApJ...597..998L}. 
}
\label{knotPolariz}
\end{figure}

We start with the case where emission come only from the shock surface. 
Figure~\ref{knotPolariz} shows the degree of polarization for the total 
flux coming from the shock with $n=2$ and proper emissivity scaling as $R^{-2}$.     
(the results for $n=4 \theta$ look very similar.)
One can see that for $0<\sigma_1<1$ the degree of polarization varies
only slightly. For the observed value of $p=2.6$, we obtain
$\Pi\approx50\%$. For magnetization $\sigma_1\geq 1$, 
the polarization signal nearly coincides with that for a spherical
shock \citep{2003ApJ...597..998L} -- in this case the flow
deflection angle $\Delta \delta $ is small and its speed is highly 
relativistic.

Thus, our model predicts high polarization of the knot emission in 
agreement the observations, but the predicted value is still somewhat lower 
than the observed one of $\Pi\sim 60\%$ \citep{2013MNRAS.433.2564M}. 
In order to understand the reason we carried out additional polarization 
calculations.

\begin{figure}
\centering
\includegraphics[width=\linewidth]{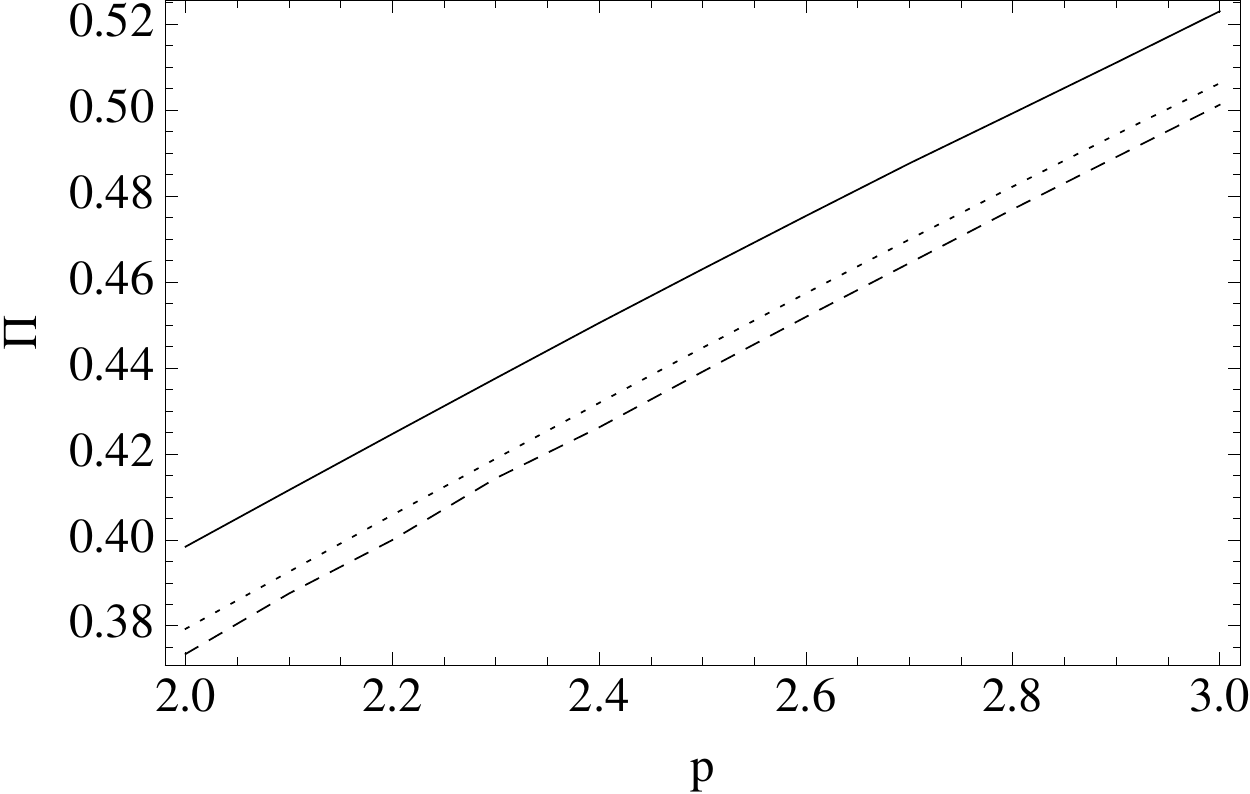}
\caption{Polarization fraction in the model with finite width of the emitting
  layer ( $\sigma_1=0$ ). The width is scaled with the local shock
  radius: $10\%$ - dashed, $20\%$ - dotted
  and $50\%$ solid.  Even for a unrealistically
  wide emitting region of $50\%$, the observed polarization degree of
  $\sim60\%$ still cannot be reproduced.  }
\label{fig:knotPolarizThick}
\end{figure}

To check if the finite thickness of the emitting layer can effect 
the polarization, we carried out calculations with the same volume 
emissivity model as in Sec.\ref{layer}.
The results are presented in Figure~\ref{fig:knotPolarizThick}, which 
shows that the degree of polarization remains largely unchanged -- 
the changes are quite small -- when the thickness 
is increased from $10\%$ to $50\%$ of the local shock radius, the
polarization increases by merely $2\%$.  This leads us to conclude
that a finite extent of the emitting region alone is unlikely to explain 
the high observed degree of polarization.

\begin{figure*}
\centering
\includegraphics[width=0.49\linewidth]{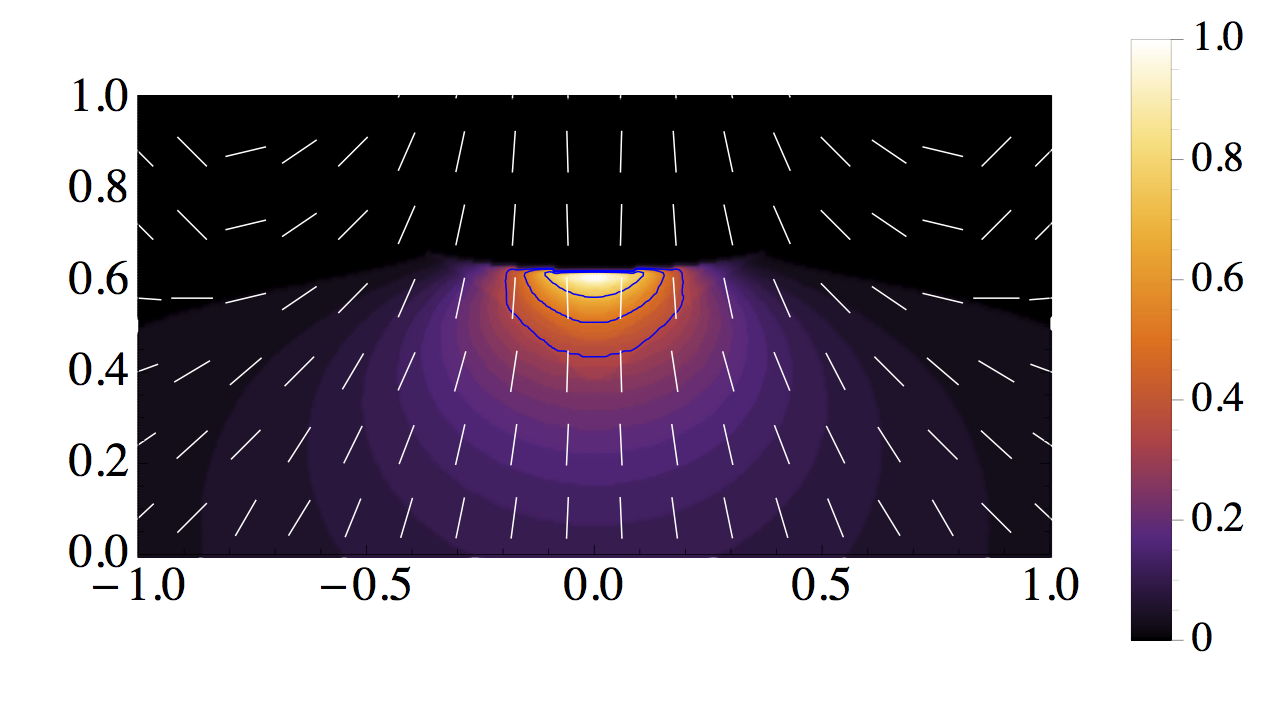}
\includegraphics[width=0.49\linewidth]{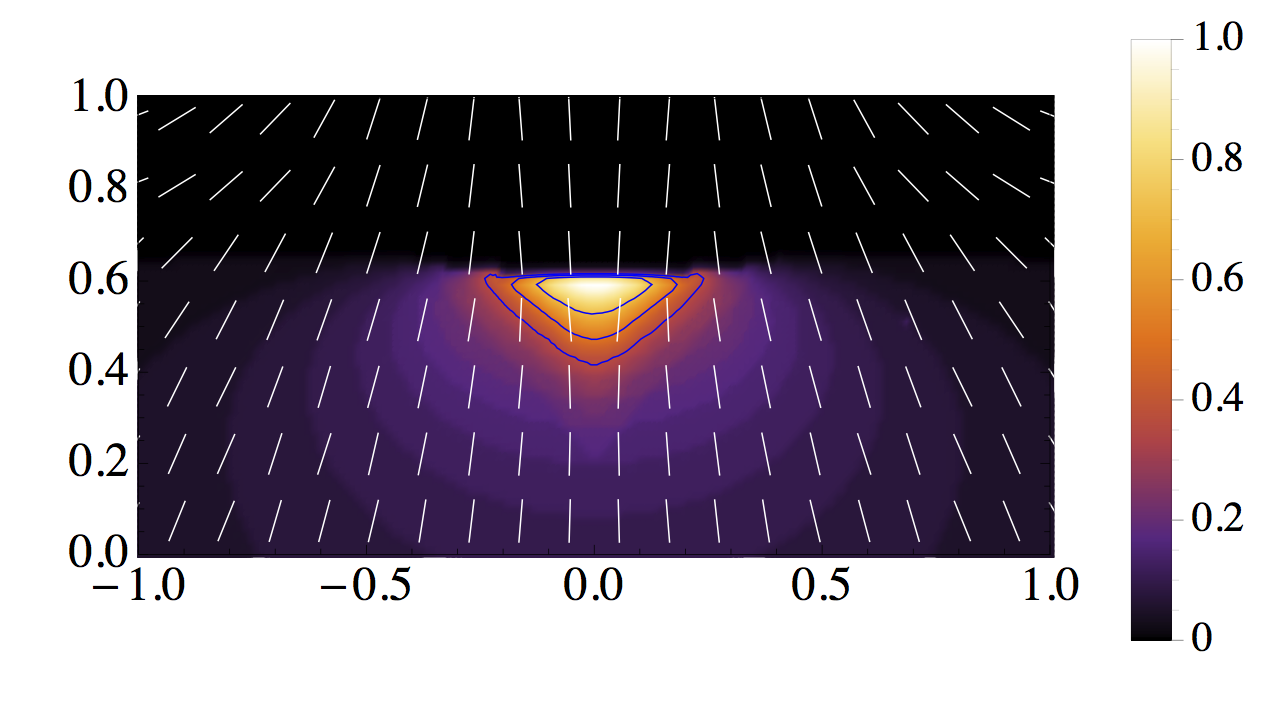}
\caption{Image of the knot ($\sigma_1=0$) and its polarization vectors
  (E-field) for $f(\theta)=\sin^2\theta$, (\textit{left}) and
  $f(\theta)=\sin^4\theta$, (\textit{right}).
At each location the polarization degree corresponds to the
theoretical maximum for the synchrotron emission.  The rotation of
polarization vectors across the image results in depolarization of the
integral emission. 
The contours start from 0.9 of the peak value
and decrease by the factor of $\sqrt{2}$ thereafter.
}
\label{EVPA}
\end{figure*}

In the observations of \cite{2013MNRAS.433.2564M}, the knot polarization
was measured in a very localized area with an aperture radius of
$0.15''$. However according to the data by \citet{2015arXiv150404613R}, 
the transverse FWHM size of the knot is $\approx 0.32\arcsec$ 
and FWRMS size is $\approx 0.56\arcsec$ and hence the aperture used in 
\citet{2013MNRAS.433.2564M} captures only the bright inner part of the knot.  
This can be significant because the depolarization of total flux in our calculations
is caused by the gradual rotation of the polarization vector across the 
knot, which is illustrated in Figure~\ref{EVPA}. 
Thus, a smaller area of integration would give a higher polarization degree. 

In order to investigate this effect we carried out additional calculations 
where the integration over the azimuthal angle was limited to the interval 
$(-\phi_b,+\phi_b)$. To determine a reasonable range for $\phi_b$, we 
recall that the knot emissivity decreases by the factor of two from its peak 
value for streamlines making angle $\alpha_h\approx 0.33/\Gamma_2\approx 0.14$ 
to the one leading to the peak (see Section~\ref{tr-size}). 

\begin{figure}
\centering
\includegraphics[width=0.99\linewidth]{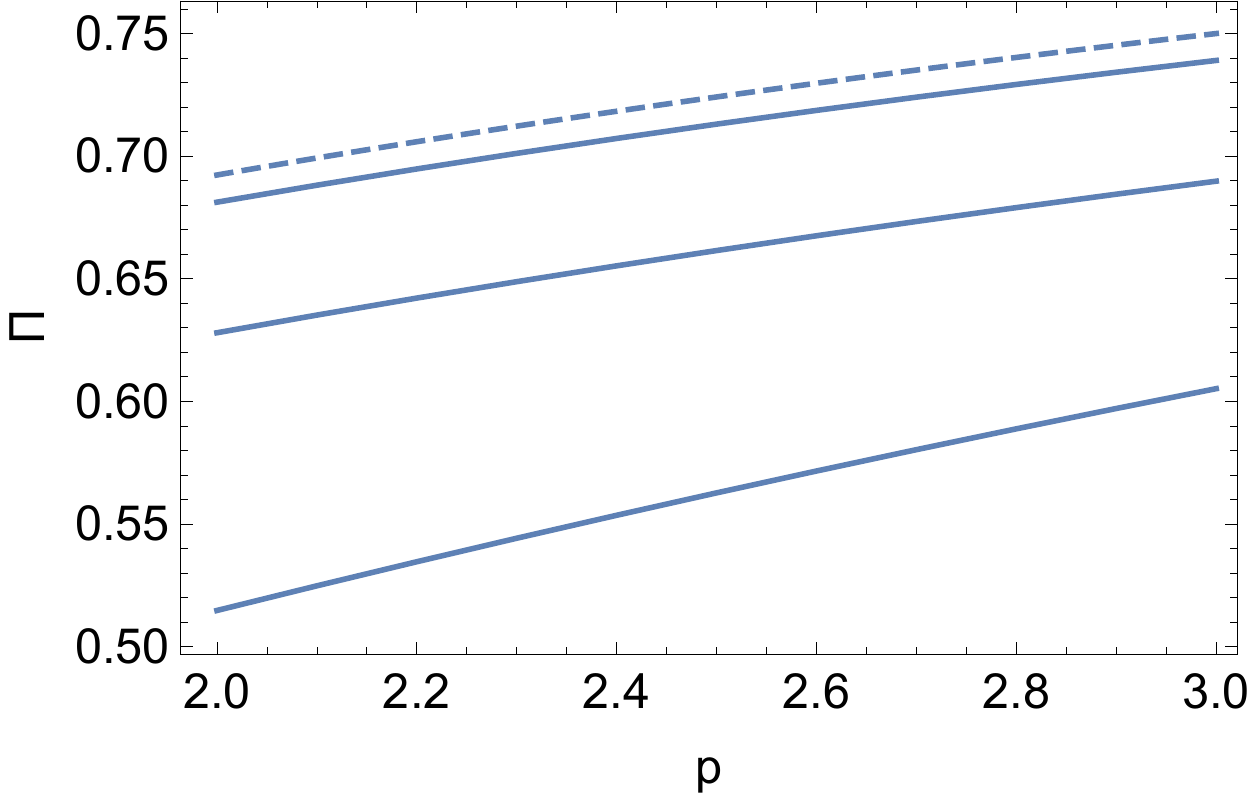}
\caption{Polarization signal coming from the area of the shock limited
  by $|\phi| \leq 0.1, \, 0.25, \, 0.5$. (At the peak of emission the
  abscissa of $ 0.15 $ arcseconds corresponds to $ \phi\approx 0.1$.)
  The dashed line is the theoretical maximum for the synchrotron
  emission in uniform magnetic field.}
\label{knotPolariz11}
\end{figure}

In Fig.~\ref{knotPolariz11}, we show the results
of integration for $\phi_b = 0.1$, 0.25 and  0.5. One can see that for all these 
value the polarization is significantly higher compared to what we obtained 
previously. In fact, for  $\phi_b=0.1$ the flux polarization degree almost coincides 
with the theoretical maximum in uniform magnetic field. Deviations from the exact 
axial symmetry of our model will naturally reduce the polarization degree.

\section{Other  properties}

\subsection{Energetics}

Let us estimate the energy flux intercepted by the region producing the wind and 
compare it with the observed luminosity. To calculate the solid angle occupied 
by the knot, we recall that for $\sigma_1\ll1$  the part of the flow which 
contributes to the knot emission occupies $\Delta \theta \approx 0.3$ with a   
$\Delta\phi \approx 0.66/\Gamma_2 \approx 0.3$ in the case $\sigma_1\ll1$ 
(see Sec.\ref{EKS}). In the same limit, the deflection angle 
$\Delta\delta \approx \delta_1 (1-\chi) \approx \theta_k/3$, where $\theta_k$
is the coordinate of knot center. This gives 
$\theta_k = \theta_{ob}-\Delta\delta \approx 3\theta_{ob}/4 \approx \pi/4$ and 
$\delta_1 \approx \pi/8$. 
The wind luminosity per unit solid
angle $L(\theta) = {3 \over 8 \pi} L_w \sin ^2 \theta$, where $L_w$ is the total
wind power. Thus, the energy passing through the knot is

\be
L_{knot} \approx 4 \times 10^{-3}  L_w  = 2 \times 10^{36} {\rm erg/s} \,,
\ee
where $L_w = 5\times 10^{38}{\rm erg/s}$ is the current spin-down power of the Crab
pulsar.
Given the observed spectrum of the nebula, most of this energy is carried out by 
the electrons emitting in the optical band. 

The observations give the isotropic optical-IR luminosity of 
the knot of $1.3\times10^{33} {\rm erg s}^{-1}$  \citep{2015arXiv150404613R}. Given 
the Doppler beaming angle of $1/\Gamma_2$ with $\Gamma_2\approx\delta_1^{-1} =  2.5$ 
the actual total knot luminosity is $L_{ob} \approx 1.6\times10^{31} {\rm erg/s}$, 
implying the radiative efficiency of $f_{rad} \approx 10^{-5}$.  

The synchrotron life-time of optical electrons is 

\be
   t_{syn} \approx 2\times10^5 B_{-3}^{-3/2} \mbox{days}\,, 
\ee
where $B_{-3}$ is the magnetic field in mGauss. Taking the knot size along streamlines of  
$t_{lc}\approx 10\,$ light days as a reasonable estimate, the knot crossing time in the fluid frame 
will be $t_{lc} \approx 4$ days. Hence, the theoretical radiative efficiency of the 
knot $f_{rad} = t_{lc}/t_{syn} \approx 2\times10^{-5} B_{-3}^{3/2}$, which is consistent with
the observations.

For comparison, the total isotropic luminosity of bright wisps within $10\arcsec$ from
the pulsar is about ten time that of the inner knot, $\approx 10^{34}{\rm erg/s}$ 
\citep{1995ApJ...448..240H}\footnote{We used the wisp length of $3\arcsec$, as stated 
in \citep{1995ApJ...448..240H} for the ``thin wisp'', to calculate 
the ratio of the isotropic wisps luminosity to that of the inner knot.}. 
Their proper motion indicates velocities $v\approx 0.6c$. Hence the beaming angle in the 
$\theta$ direction is $\Delta\theta \approx 0.6$ and in the $\phi$ direction 
$\Delta\phi \approx 0.5$, slightly smaller due to the anisotropy of the synchrotron 
emissivity in uniform magnetic field.  The corresponding solid angle is about unity and 
hence the actual wisps luminosity is $L_{ob}\approx (1/4\pi)\times10^{34}{\rm erg/s}$. 
The luminosity emitted in all directions will be higher on average by $\pi/\Delta\phi$, 
yielding $L_{wisps}\approx 5\times10^{33}{\rm erg/s}$.  
According to the MHD theory,
the wisps are arc-like structures of enhanced magnetic field advected in the equatorial
direction \citep{2009MNRAS.400.1241C,2014MNRAS.438..278P}. It is in this direction, 
where most of the pulsar wind power is transferred. Hence the radiation efficiency of 
the wisp region 

\be
    f_{rad}=\frac{L_{wisps}}{L_{w}}\approx 10^{-5} \,, 
\ee    
which is similar to what we found for the knot\footnote{
The maximal spectral power in the Crab Nebula comes out in
  UV-soft X-rays, where the radiative times scales of leptons is
  roughly comparable to the age of the nebula. Most of this emission
  comes from the old volume-filling population of particles, and not
  from the freshly accelerated ones close to the termination shock.}. 

\subsection{Variability}

\cite{2013MNRAS.433.2564M,2015arXiv150404613R} discuss the variability
of the position, size and the luminosity of the inner knot - the
position fluctuates relative to the mean by approximately 10\% on the
time scales of month(s). At the same time the overall size of knot
correlates with the distance from the pulsar, while the luminosity
anticorrelates with it.  

The numerical simulations by \cite{2009MNRAS.400.1241C,2014MNRAS.438..278P} 
show that inner region of PWN is highly dynamic and the shock surface 
is constantly changing as the result. When the external pressure drops 
the shock expands and when the pressure increases the shock recedes.  
The emission of wisps is one effect of this variability observed in the 
synthetic synchrotron images obtained in the simulations. 
The other one is the unsteady behavior of the inner knot, whose position and 
brightness change in time. In fact, \cite{2014MNRAS.438..278P} reported an
anti-correlation of their synthetic knot luminosity with its projected
distance from the pulsar. 

In order to understand these results, let us consider the simplest model 
of the shock variability, where the shock shape is preserved but its 
length scale $R_0$ fluctuates. In this case, the downstream emissivity is
the same function of $\theta$ and $\phi$ up to a factor depending on 
$R_0$. This means that the ratio of the knot size to its separation from 
the pulsar remains unchanged (see Sec.\ref{EKS}), which is in good agreement 
with the HST observations, which give $\psi_\perp \propto \psi_k^{0.8\pm0.13}$.  
    
Regarding the total flux from the knot, we note the emissivity 
$\epsilon_\nu \propto n'_2 B_2^{'(p+1)/2}$, where $n'$ the number density 
of emitting particles. The total flux of the knot $F_\nu \propto \epsilon_\nu A$, 
where $A$ is the knot area. Since, $n'_2\propto R_0^{-2}$, $A\propto R_0^2$, 
$B'_2\propto R_0^{-1}$ and $\psi_k\propto R_0$ we obtain 

\be 
    F_\nu \propto \psi_k^s \quad\mbox{where}\quad s= -(p+1)/2\,;
\label{flux-separation}
\ee
the same result as stated in \citet{2015arXiv150404613R}. Thus, the shock model 
is consistent with the observed anti-correlation. 
For the observed spectral index $p=2.6$, Eq. (\ref{flux-separation}) gives $s=1.8$, 
whereas the HST data suggest a somewhat larger value $s=2.39\pm0.37$ 
\citep{2015arXiv150404613R}. In reality, the shock variability may not be 
shape-preserving, in which case the variablity of its observed emissivity will be 
more complicated.    
      
The results of computer simulations show that the shock variability is 
more complicated, with the shock shape changing as well in response to 
the external perturbations on the scale below $R_0$ \citep{2014MNRAS.438..278P}.  
Thus, the predictions based on the model of uniform scaling should be considered 
as rather approximate.

Since in the MHD theory both the wisp production and the knot
variability are related to the variations of the shock geometry, one
would expect approximately the same time-scale for both these
phenomena\footnote{Knot variability may occur on shorter scales  
\citep{2012MNRAS.422.3118L}.}. Although the
available observational data do not cover a sufficiently long period
of time, they indicate that this may be the case  
\citep{2002ApJ...577L..49H,2013MNRAS.433.2564M,2015arXiv150404613R}.

\subsection{Connection to Crab's gamma-ray flares}

The discovery of flares from the Crab Nebula
\citep{2011Sci...331..736T,2011Sci...331..739A} challenges our
understanding of particle acceleration in PWNe and possibly in
other high energy astrophysical sources.


The short life-time of gamma-ray emitting electrons means that if they are
accelerated at the termination shock then the gamma-ray emitting
region is a thin layer above the shock where the flow Lorentz factor is still high 
and hence its emission is subject to the Doppler beaming. 
\citet{2011MNRAS.414.2017K} used this to argue that
most of the observed gamma-ray emission of the Crab Nebula may come
from the inner knot.  They and
\citet{2012MNRAS.422.3118L} also speculated that the gamma-ray flares
of the Crab Nebula may come from the knot as well and proposed to look
for correlations between the knot's optical emission and the gamma-ray
emission from the nebula.  The relativistic post-shock flow may help to
explain the peak frequency of flares exceeding the radiation reaction limit
of  $\approx 100\,$MeV \citep{2010MNRAS.405.1809L,2011MNRAS.414.2017K}.  
Moreover, a blob moving through the knot of $\approx 10$ light days length would be
observed for the time smaller by $ 2 \Gamma_2^2$, which can explain the short 
timescales of the Crab's flares.

The observed cutoff of the synchrotron spectrum of the Crab Nebula at
$\sim 100\,$MeV in the persistent Crab Nebula emission and especially
during the flares, when the cutoff energy approached even higher value
of $\sim 400\,$ MeV, is in conflict with slow stochastic acceleration mechanisms
\citep{2010MNRAS.405.1809L,2012MNRAS.426.1374C}. Alternatively, the flares 
may result from linear particle acceleration during explosive relativistic 
magnetic reconnection 
\citep[\eg][]{LyutikovUzdensky,2005MNRAS.358..113L,2010MNRAS.405.1809L,2014PhPl...21e6501C}.
Fast and efficient particle acceleration during the reconnection requires
highly magnetized plasma, $\sigma \geq 1$, in the flare producing
region \citep[\eg][]{2012MNRAS.427.1497L}. However, our results show 
that the knot plasma cannot be that highly magnetized. Moreover, the magnetic
field in this region is still expected to be very regular (after the
dissipation of the small-scale magnetic stripes), namely azimuthal of
the same orientation \citep{2014MNRAS.438..278P}, and hence lacking
current sheets required for the reconnection.
Finally, the coordinated programs of optical observations
did not reveal anything unusual about the inner knot emission during
the gamma-ray flares \citep{2013ApJ...765...56W,2015arXiv150404613R}.
Thus, we have to admit that the Crab flares are unlikely to originate 
from its inner knot. 

If flares do not come the termination shock then they are certainly not 
connected to the shock particle acceleration mechanism. An explosive magnetic reconnection 
seems to be the only realistic alternative. A favorable location for such 
reconnections would have high magnetization parameter $\sigma =B^2/w >1$, where 
$w$ is the relativistic enthalpy, as this would ensure the relativistic 
Alfv\'en speed. In addition, its magnetic field should be somewhat disordered 
so that thin current sheets may develop. In
pulsar wind nebulae, such plasma is expected to exist in the polar
region downstream of the termination shock, which is fed by the
unstriped section of the pulsar wind
\citep{2012MNRAS.427.1497L,2013MNRAS.428.2459K,2014MNRAS.438..278P}.

The current observations do not rule out yet that at energies below
100~MeV the synchrotron gamma-ray emission between flares is coming
from the knot. If so, a slow variability of the persistent gamma-ray emission 
at these energies, on the time-scale of wisp production, is expected. 
Additional studies are required to clarify this issue.

\section{Comparison with other studies}

Almost simultaneously with our manuscript, the results of an independent 
study by \cite{2015MNRAS.454.2754Y} have become publicly available (Some of their 
results have been outlined already in \citep{2015arXiv150404613R}.). 
We agree in the conclusion that the observed knot parameters rule out high 
magnetization of the post-shock plasma. However, they could not reach a definitive 
conclusion on the acceptability of the shock model even in the low magnetisation regime, 
pointing to a number of difficulties. The main of them concern the transverse size of the knot, 
its shape and polarization. 

For the transverse size, \cite{2015MNRAS.454.2754Y} claimed that in the basic shock model 
it exceeds the distance to the pulsar at least by the factor of 2.8, in conflict with the 
observations. 
However, this estimate is based on the assumption that the emissivity drops by the factor of 
two at an angle $\alpha_h=1/\Gamma_2$ to the velocity vector. In reality, the combination 
of the Doppler beaming with the anisotropy of the proper synchrotron emissivity in uniform 
magnetic field leads to a much smaller angle (see Section~\ref{tr-size}). Curiously, 
their synthetic image in Figure~3b  shows a much more compact knot, well in line with our 
results.  

For the polarization of the integral shock emission, they obtained a value which is lower 
compared to that of the inner knot as obtained by \cite{2013MNRAS.433.2564M}. In fact, this result 
agrees with our calculations, when the flux integration is carried out over the whole shock 
surface. However, in the observations, the polarization is measured only for the bright core of 
the knot (area with an aperture radius of $0.15"$). We have demonstrated that smaller integration 
area leads to higher polarization degree, allowing a much better fit. 

Finally, \cite{2015MNRAS.454.2754Y} pointed out that the shock model cannot
reproduce the ``smily'' shape of the knot, claimed in \citet{2015arXiv150404613R}, 
but yields images more reminicent of a ``frown''. 
This conclusion is base on the model where the emission comes only from the shock surface, 
which also leads to a very sharp brightness drop at the distant (relative to the puslar) 
edge of the knot. We have shown that in models with finite thickness of 
the emitting layer, these features do not survive and the frown can easily turn into 
a smile or even a pout. In fact, the experimentation with different geometries of 
streamlines in the emitting zone by \cite{2015MNRAS.454.2754Y} also show rather strong 
distortions of synthetic images. To address such details more advance models, based on 
computer simulations, are required.

\section{Conclusion}

In this paper, we have further explored the model of the Crab Nebula inner 
knot as a Doppler-boosted emission from the termination shock of the pulsar wind.
This model successfully explains a number of its observed properties: 

\begin{description} 

\item{\it Location:} The knot is located on the same side of the pulsar as the Crab jet,   
along the symmetry axis of the inner nebula, and on the opposite side as the brighter 
section of the Crab torus. This is a direct consequence 
of the termination shock geometry and the Doppler-boosting.  

\item{\it Size:} The knot size is comparable to its separation from the pulsar. 
This also follows from the shock geometry and the Doppler-beaming. The anisotropy 
of the proper synchrotron emissivity, which vanishes along the magnetic field 
direction in combination with the relativistic aberration of light is another 
significant factor. Only models with low magnetization of the post-shock flow,
with the effective magnetization parameter of the wind $\sigma_1 < 1$ agree with 
the observations. 

\item{\it Elongation:} The knot is elongated in the direction perpendicular to the symmetry 
axis. This is because the knot emission comes from the region where the shock surface is 
almost parallel to the line of sight.

\item{\it Polarization:} The knot polarization degree is high, and 
the electric vector is aligned with the symmetry axis. This come due to the fact that 
the post-shock magnetic field is highly ordered in the vicinity of the termination shock 
and azimuthal. In the model, the relativistic aberration of light leads to a noticeable 
rotation of the polarization vector along the knot and this prediction could be tested in 
future polarization observations. Accordingly, the polarization degree of the integral  
knot emission depends on the integration area - the bigger the area the smaller the 
degree is.        

\item{\it Luminosity:} Taking into account Doppler beaming, the observed radiative 
efficiency of the inner knot is consistent with efficient particle acceleration at the 
termination shock and the knot's magnetic field of one milli-Gauss strength, 
which is a reasonable value for the inner Crab Nebula. 

\item{\it Variability:} The knot flux is anti-correlated with its separation from the pulsar. 
In the numerical simulations, the termination shock is found to be highly unsteady, changing its
size and shape. As the shock moves away from the pulsar, so does the knot region, which leads 
to lower magnetic field and hence lower emissivity. Another outcome of the shock variability
in the MHD simulations is the emission of wisps and hence one expects both the processes 
to occur on the same time-scale, which is consistent with the observations.        
  
\end{description} 

In many cases, the agreement with the observed properties of the Crab's inner knot falls short
of a perfect fit.  Given the uncertanties in the shape of the termination shock, proper emissity 
of the shocked plasma and the post-shock flow which are present in the model it would be naive to 
expect more. Further investigations of the models, involving advanced numerical simulations, 
are needed to achive this.    

Our results may have a number of important implications to the astrophysics of 
relativistic plasma in general and that of PWN in particular. They show that 
the termination shock of the relativistic wind from the Crab pulsar is a 
reality and that this shock is a location of efficient particle acceleration.      
The strong Doppler-beaming of the emission from the shock explains why this 
shock has been so elusive. Only the emission from a small patch on the shock 
surface, the inner knot, is strongly Doppler-boosted and hence prominent. 
For most of the shock, its emission is beamed away from the Earth and hence 
difficult to observe.

The shock model of the inner knot allows us to constrain the parameters of 
the wind from the Crab pulsar. Taken directly, the model requires the wind to 
be particle-dominated, $\sigma_1 < 1$ , at least at the polar latitudes of 
$40^\circ-60^\circ$. However, in the case of a striped wind, its termination 
shock can mimic that of a low $\sigma$ flow even when the actual wind magnetization 
is extremely high \citep{2003MNRAS.345..153L}. In this context, the   
magnetic inclination angle of the Crab pulsar should be above $45^\circ$, 
which means that  most of the Poynting flux of the Crab wind is converted into 
particles, if not in the wind itself then at its termination shock 
\citep{2013MNRAS.428.2459K}.  This is in agreement with the results of numerical 
simulations, which can reproduce the observed properties of the inner Crab 
Nebula extremely well in models with moderate wind magnetization 
\citep{2014MNRAS.438..278P}. However, the polar region of a pulsar 
wind is free of stripes and can still inject highly magnetized plasma into 
its PWN.  

The fact that during the gamma-ray flares of the Crab nebula the inner knot 
does not show any noticeable activity suggests that the flares occur 
somewhere else. This is consistent with the fact that any stochastic 
acceleration mechanism is too slow to compete with radiative losses and 
deliver electrons capable of emitting synchrotron photons of $100\,$MeV energy.  
Our conclusion that the inner-knot plasma is not highly magnetized also
disfavors the knot as a site of explosive relativistic magnetic reconnection.   
To proceed really fast, the magnetic reconnection has to occur in 
magnetically-dominated plasma 
\citep{2010MNRAS.405.1809L,2012MNRAS.426.1374C,2014PhPl...21e6501C,2012MNRAS.427.1497L,2014ApJ...783L..21S}. 
The inner polar region of the Crab nebula is the only location where such conditions 
can be met.

\section{Acknowledgments}
    
We would like to thank Roger Blandford for stimulating discussions of
the issue.  We also thank Paul Moran for information on the details of the 
polarization observations and Lorenzo Sironi for comments on the manuscript. 

This work had been supported by NASA grant NNX12AF92G  and  NSF  grant AST-1306672.
SSK and OP are supported by STFC under the standard grant ST/I001816/1.

\bibliographystyle{mn2e}

\bibliography{/Users/maxim/Home/Research/BibTex}

\appendix
\onecolumn
\section{Oblique  relativistic MHD shocks}
\label{oblique}

In the shock frame, the fluxes of energy, momentum, rest mass, and
magnetic field are continuous across the shock
\begin{equation}
\label{s1}
(w+B^2)\gamma^2\beta_x = \mbox{const},
\end{equation}
\begin{equation}
(w+B^2)\gamma^2\beta_x\beta_x +p +\frac{B^2}{2} = \mbox{const},
\label{s2}
\end{equation}
\begin{equation}
(w+B^2)\gamma^2\beta_x\beta_y = \mbox{const},
\label{s3}
\end{equation}
\begin{equation}
\rho\gamma\beta_x = \mbox{const},
\label{s4}
\end{equation}
\begin{equation}
B\gamma\beta_x = \mbox{const},
\label{s5}
\end{equation}
where $\rho$ is the rest mass density, $p$ is the gas pressure, $w=\rho c^2 + \kappa P$ is
the relativistic enthalpy, $\kappa=\Gamma/(\Gamma-1)$, where $\Gamma$ is the adiabatic
index, $B$ is the magnetic field as measured in the fluid frame,
$\beta=v/c$, and $\gamma$ is the Lorentz factor.
We select the frame where the velocity vector is in the xy-plane, the magnetic field
is parallel to the z-direction, and the shock front is parallel to the yz-plane.
In what follows we will use subscripts 1 and 2 to denote the upstream an the downstream
states respectively. 

We assume that the upstream plasma is cold, $p_1=0$, 
and ultrarelativistic, $\gamma_1 \gg 1$,  
that the shock is strong and the downstream ratio of specific heats is $\Gamma_2=4/3$.
Hence $\beta_{1x}\approx\sin\delta_1$, 
there $\delta_1$ is the angle between the velocity vector and the shock plane.  
Denote the wind energy flux in the radial direction as $F$. Then 

$$
  (w_1+B_1^2)\gamma_1^2\beta_{1x} = (F/c) \sin\delta_1\,,
$$ 
and Equations \ref{s1} and \ref{s2} read 
\begin{equation}
\label{s1a}
(w_2+B_2^2)\gamma_2^2\beta_{2x} = (F/c) \sin\delta_1,
\end{equation}
\begin{equation}
(w_2+B_2^2)\gamma_2^2\beta_{2x}\beta_{2x} +\tilde{p}_2 = (F/c) \sin^2\delta_1,
\label{s2a}
\end{equation}
where $\tilde{p}_{2} = p_2 +\frac{B_2^2}{2}$ is the total pressure (Note that we ignore 
the contribution of the magnetic pressure to the upstream momentum flux). 
Combining the two one finds 
\begin{equation}
\tilde{p}_2 = (F/c) (\sin^2\delta_1-\beta_{x2}\sin\delta_1)\,. 
\label{pt2}
\end{equation}
For a strong shock, 
$$
  \beta_{2x} = \chi\beta_{1x}=\chi\sin\delta_1\,,
$$ 
where 
\begin{equation}
\chi=\frac{1+2\sigma_1 + \sqrt{16\sigma_1^2+16\sigma_1+1}}{6(1+\sigma_1)}
\label{os8}
\end{equation}
\citep{2011MNRAS.414.2017K}. Hence,  
\begin{equation}
\tilde{p}_2 = (1-\chi) (F/c) \sin^2\delta_1 \,. 
\label{pt2a}
\end{equation}
$\chi$ is a monotonically decreasing function of $\sigma_1$. 
For $\sigma_1=0$, one has $\chi=1/3$ and 
\begin{equation}
\tilde{p}_2 = \frac{2}{3}\frac{F}{c} \sin^2\delta_1\,  
\end{equation}
which is the same as derived in \citet{2014MNRAS.438..278P}. For $\sigma_1\gg1$, one has 
$\chi\simeq 1-1/2\sigma_1$ and 
\begin{equation}
\tilde{p}_2 = \frac{1}{2\sigma_1}\frac{F}{c} \sin^2\delta_1\,.  
\end{equation}

\section{Emissivity calculations.}
\label{app-emission}

Let us introduce Cartesian coordinates centered on the pulsar with the z axis aligned with 
its rotational axis and the line of sight parallel to the XOZ plane. In the corresponding 
basis, the radius vector of a point on the shock surface is    
${\bf R}_s = R_s (\sin\theta\cos\phi,\sin\theta\sin\phi,\cos\theta)$. 
The orthogonal projection of this vector into the plane of the sky is \be
{\bf R}_{s,\perp} =  {\bf R_s} - ({\bf R_s}\cdot{\bf n}_{ob}) {\bf n}_{ob} \,,
\label{rad-vec-pr} 
\ee
where ${\bf n}_{ob}$ is a unit vector along the line of sight. 
In the plane of the sky, we introduce the angular polar coordinates $\{\psi_{ob},\phi_{ob}\}$  with 
the origin at the pulsar image and the reference direction given by the orthogonal projection of the 
rotational axis (see Fig.~\ref{geom1}). Given Eq. (\ref{rad-vec-pr}), the projection of the 
shock point has the coordinates
\ba &&
\psi_{ob} = (R_s/D)   \sin \Theta \,,
\nn && 
\sin \phi_{ob} ={ {\bf R}_{s,\perp} \cdot {\bf n_y} \over | {\bf R}_{s,\perp}|}=
  {\sin \phi \sin \theta \over \sin\Theta}\,,
\nn &&
\cos \Theta = \cos \theta \cos \theta_{ob} + \cos \phi  \sin  \theta \sin \theta_{ob} 
\label{1}
\ea
and ${\bf n}_y = \{0,1,0\}$ is the unit vector along the y axis.

For any proper emissivity, the relativistic Doppler and aberration of light effects ensure
that the observed synchrotron emissivity
\be
\epsilon_\nu \propto {\cal D}^{2+(p-1)/2} | \sin \mu'|^{(p+1)/2},
\ee
where $p$ is the particle spectral index and $\mu'=\pi/2-\alpha'$ is the angle between the 
magnetic field and the line of sight in the fluid frame \citep[\eg][]{2003ApJ...597..998L}.

At the shock, the post-shock velocity direction is given by the unit vector 

\be
{\bf n}_f= \{\cos \phi \sin \theta_f, \sin \phi \sin \theta_f, \cos \theta_f\} \,,
\ee 
where $\theta_f= \theta + \Delta\delta$. Hence the Doppler factor 

\be
{\cal D} = \left(\Gamma_2 (1 - v_2 ( {\bf n}_f \cdot {\bf n}_{ob})) \right)^{-1}\,.
\ee
In the fluid frame, the direction vector of the line of sight is 

\be
{\bf n'}_{ob} =
\frac{ {\bf n}_{ob} + \Gamma_2 {\bf v}_2 \left( {\Gamma_2\over \Gamma_2+1} 
({\bf n}_{ob} \cdot {\bf v}_2) -1\right)} 
{ \Gamma_2 \left(1- ({\bf n}_{ob} \cdot {\bf v}_2)\right)}
\ee
\citep[Eq. C9 in][]{2003ApJ...597..998L}  and since the magnetic field 
is purely azimuthal\be\cos \mu' = {{\bf n}_\phi \cdot {\bf n'}_{ob} }\,.
\ee

The unit electric polarization vector (EPV) of synchrotron emission can be found as 

\be
   {\bf e} = \frac{\ort{ob}\times [\ort{\phi} + \ort{ob}\times({\bf v}_2\times\ort{\phi})]}
         {\sqrt{(1-\ort{ob}\cdot{\bf v}_2)^2-(\ort{\phi}\cdot\ort{ob})^2/\Gamma_2^2}}  
\ee  
\citep{2003ApJ...597..998L}.
The angle $\tilde\chi$ between this vector and the symmetry axis in the plane of the sky 
is given by

\be
     \cos\tilde\chi = {\bf e}\cdot(\ort{ob}\times\ort{y})\,, \quad 
     \sin\tilde\chi = -({\bf e}\cdot\ort{y})\,. 
\ee
According to these equations, it increases in the anti-clockwise direction.
Fig.~\ref{EVPA} shows the distribution of the this vector over the synthetic image 
of the knot in the model with $f(\theta)=\sin^2\theta$ and $\sigma_1=0$. One can see that the 
vector is rotating across the knot. At each point, the polarization degree is maximal but the 
polarization of integral emission will be lower due to this rotation.

Due to the mirror symmetry of the image the Stokes parameter ${\bar{U}}$ integrates 
to zero, and the polarization fraction of integral emission is

\be
\Pi ={ |{\bar{Q}} | \over {\bar{I}}} = \frac{p+1}{p+7/3} 
 {\int \epsilon(\theta,\phi) \cos 2{\tilde\chi} dV \over 
\int \epsilon(\theta,\phi) dV}  \,.
\label{Pi}
\ee
In models where the emission comes from the shock surface, the emissivity 
includes the delta-function $\delta(R-R_s(\theta))$, where $R_s(\theta)$ is the 
shock radius.

\end{document}